\documentclass[10pt]{article}
\pdfoutput=1
\usepackage{jheppub}
\usepackage{amsmath,amssymb,amsfonts,mathbbol,graphicx,slashed,color,amsthm, mathtools, upgreek, enumerate, tensor}
\usepackage{subcaption}
\usepackage{setspace}
\usepackage[export]{adjustbox}
\usepackage{arydshln}
\usepackage[dvipsnames]{xcolor}
\usepackage{physics}
\usepackage[normalem]{ulem}
\usepackage{comment}
\usepackage{dsfont}

\usepackage{tikz}
\usetikzlibrary{shapes.geometric}
\usetikzlibrary{decorations.markings}
\usetikzlibrary{decorations.pathmorphing}

\usepackage{pgfplots}
\pgfplotsset{compat=newest}
\usepgfplotslibrary{fillbetween}

\allowdisplaybreaks

\colorlet{darkblue}{blue!70!black}
\colorlet{darkgreen}{green!50!black}
\colorlet{darkred}{red!50!black}

\newcommand{\dg}[1]{\textcolor{darkgreen}{#1}}

\def\bea{\begin{eqnarray}}
\def\eea{\end{eqnarray}}
\def\be{\begin{equation}}
\def\ee{\end{equation}}

\title{
On the Spread of Entanglement at Finite Cutoff
}

 \author[a]{Evan Coleman,\footnote{Now at MIT Climate \& Sustainability Consortium.}}
 \author[a,b]{Ronak M Soni,}
 \author[a]{and Sungyeon Yang}
 \affiliation[a]{Stanford Institute for Theoretical Physics, 382 Via Pueblo, Stanford CA 94305}
 \affiliation[b]{Department of Applied  Mathematics and Theoretical Physics,  University of Cambridge, Wilberforce Road, Cambridge, CB3 0WA, UK}
 \vspace{5mm}

\vspace{1cm}

\emailAdd{evanacoleman@gmail.com}
\emailAdd{rs2194@damtp.cam.ac.uk}
\emailAdd{syang61@stanford.edu}

\abstract{
  We study how entanglement spreads in the boundary duals of finite-cutoff three-dimensional theories with positive, negative and zero cosmological constant, the $T \bar{T} + \Lambda_{2}$ two-dimensional theories.
We first study the Hawking-Page transition in all three cases, and find that there is a transition in all three scenarios at the temperature where the lengths of the two cycles of the torus are the same.
We then study the entanglement entropy in the thermofield double states above the Hawking-Page transition, of regions symmetrically placed on the two boundaries.
We consider the case where the region is one interval on each side, and the case where it is two intervals on each side.
We give an entanglement tsunami interpretation of the time-evolution of the entanglement entropies.
}

\begin{document}

\maketitle
\parskip=10pt

\section{Introduction}
\label{sec:intro}

An important question about many-body systems is the pattern of entanglement spreading in a high-energy state.
In integrable systems, this pattern is determined by free-streaming quasi-particles \cite{Calabrese:2009qy}, but this doesn't generalise to arbitrary systems, as shown in e.g. \cite{Asplund:2015eha}.
In highly chaotic systems, there are other coarse-grained heuristics, in terms of a minimal membrane \cite{Nahum:2016muy} or, in the holographic case, an entanglement tsunami \cite{Liu:2013iza,Leichenauer:2015xra}.

The set-up is as follows.
Consider a thermofield double state of two CFTs dual to a two-sided eternal black hole.
Take the region $A$ to be a union of intervals on each side, placed so that there is a $\mathds{Z}_{2}$ symmetry between the two exteriors.
Then, the upper bound for the time-evolution of the entanglement entropy of $A$ can be calculated as follows.
We assume here that all time and length scales are much bigger than the thermal length scale $\beta$.

Imagine a `tsunami' leaving the points $\partial A$, with wavefronts emanating in both directions at the speed of light.
Let us call the interior of this tsunami $E(t)$, and each connected component $E_{i} (t)$.
When two tsunami wavefronts hit each other, the connected components on each side merge.
The interior of this tsunami is a region carrying an $s_{eq}$ amount of entanglement per unit length and the exterior carries none.
Then the entanglement entropy is upper-bounded by
\begin{equation}
  S(t) \le s_{eq} \sum_{i} \min \left\{ \mathrm{vol} \left( E_{i} (t) \cap A \right), \mathrm{vol} \left( E_{i} (t) \cap A^{c} \right) \right\}.
  \label{eqn:tsunami-formula}
\end{equation}
Note that we have to perform the minimisation for each connected component of the tsunami separately.
In \cite{Leichenauer:2015xra}, it was shown for 2d holographic CFTs that this is an equality for one or two intervals on each side and satisfied for more intervals.\footnote{\cite{Leichenauer:2015xra} considered a quench state, dual to an AdS-Vaidya geometry. However, this qualitative picture applies equally well to the case described here.}

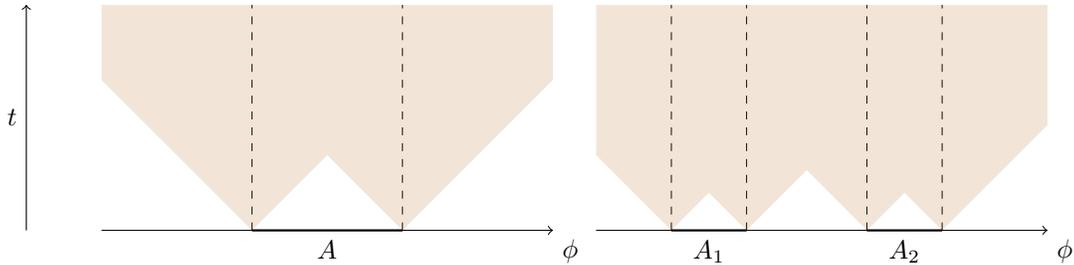
\begin{figure}[h!]
  \centering
  \begin{tikzpicture}
    \draw[->] (0,0) -- node[below right,pos=1] {$\phi$} (6,0);
    \draw[thick] (2,0) -- node[below] {$A$} (4,0);
    \draw[dashed] (2,0) -- (2,3);
    \draw[dashed] (4,0) -- (4,3);
    \draw[->] (-1,0) -- node[left] {$t$} (-1,3);
    \fill[brown,opacity=.2] (0,2) -- (2,0) -- (3,1) -- (4,0) -- (6,2) -- (6,3) -- (0,3) -- (0,2);
  \end{tikzpicture}
  \begin{tikzpicture}
    \draw[->] (0,0) -- node[below right,pos=1] {$\phi$} (6,0);
    \draw[thick] (1,0) -- node[below] {$A_{1}$} (2,0);
    \draw[thick] (3.6,0) -- node[below] {$A_{2}$} (4.6,0);
    \draw[dashed] (1,0) -- (1,3);
    \draw[dashed] (2,0) -- (2,3);
    \draw[dashed] (3.6,0) -- (3.6,3);
    \draw[dashed] (4.6,0) -- (4.6,3);
    \fill[brown,opacity=.2] (0,1) -- (1,0) -- (1.5,.5) -- (2,0) -- (2.8,.8) -- (3.6,0) -- (4.1,.5) -- (4.6,0) -- (6,1.4) -- (6,3) -- (0,3);
  \end{tikzpicture}
  \caption{The entanglement tsunami for one and two intervals on each side, respectively. Only one of the two CFTs is pictured here. The brown region is the entanglement tsunami.}
  \label{fig:tsunami}
\end{figure}

In this work, we investigate the spread of entanglement for the case of one and two intervals on each side in the case of $T \bar{T} + \Lambda_{2}$-deformed holographic CFTs.
These theories are not quantum field theories at all, meaning that the heuristics mentioned above don't necessarily generalise to these cases.

The central question of this work is: how is the picture of the entanglement tsunami modified by the $T \bar{T} + \Lambda_2$ deformation?
We attempt to answer this question using the finite-cutoff holography interpretation of these theories, for theories dual to finite-cutoff $AdS_3$, $dS_3$ and also three-dimensional flat space; we review the definition of these theories and the duality in section \ref{sec:review}.

Since the behaviour above can only occur in the presence of an ER bridge connecting the two boundaries, we first have to carefully map out the Hawking-Page phase transition in all three cases.
This is what we do in section \ref{sec:saddle}.
We find the following high-temperature bulk duals in the three cases.
In AdS, the bulk is a two-sided eternal black hole as expected.
In dS, the black hole horizon is replaced by a cosmic horizon, as we showed in \cite{Coleman:2021nor}.
In flat space, we have a Rindler horizon with the two boundaries tracing out the paths of (rings of) uniformly accelerated observers on the two sides of this horizon.
It should be noted that the appearance of the cosmic and Rindler horizons instead of black hole horizons might be due to the lack of asymptotically flat or dS black holes in three dimensions, and therefore this behavior might not generalise to higher dimensions.

In the bulk holographic interpretation of the $T\bar{T}$ deformation, the proposal of a Dirichlet wall at fixed radius in a BTZ black hole background was noted to permit superluminal boundary gravitons sourced by perturbations of the cutoff surface~\cite{McGough:2016lol}. Specifically, the fluctuations of the metric under such a shift satisfy a wave equation with phase velocity $v_\lambda > c = 1$, where $\lambda$ is the $T \bar{T}$ deformation parameter.\footnote{There are two sign conventions for $\lambda$ in the literature; we take it to be of the `holographic' sign.}
In the original proposal, the authors matched this speed to a quantity in the 2D boundary theory by absorbing the deformation into the metric of the seed CFT, treating the spacetime as Gaussian-random.

In the $AdS$ TFD, we show in section \ref{sec:ads} that in the deformed theory the wavefront of the entanglement tsunami is also superluminal (for times and distances lage compared to both the inverse temperature $\beta$ and the deformation length-scale $\sqrt{\lambda}$), with a speed that agrees with the one found for boundary gravitons in \cite{McGough:2016lol}.
Furthermore, we show that this superluminal wavefront is equivalent to a precisely luminal wavefront on the ``base space'' where the effects of the $T\bar{T}$ deformation have not been absorbed into the clocks and rods of the theory. This demonstrates how the superluminality arises from the coupling to a topological gravity theory.

We also extend these considerations to theories deformed by $T \bar{T} + \Lambda_2$, dual to finite-cutoff versions of flat and dS${}_{3}$ spaces, in sections \ref{sec:dS_EV} and \ref{sec:HPeta0} respectively.
We are able to give entanglement tsunami interpretations in both these cases.
Unlike in the AdS case, however, the wavefront of the tsunami never becomes approximately constant speed in these cases.

We end with a short discussion in section \ref{sec:discussion}.

\section{Review} \label{sec:review}
A $T \bar{T} + \Lambda_2$-deformed CFT is defined by the differential equation
\begin{equation}
  \partial_{\lambda} \log Z_{\lambda} = \int \langle T \bar{T} \rangle + \frac{1 - \eta}{\lambda^{2}},
  \label{eqn:ttbar-eqn}
\end{equation}
where $Z$ is the partition function of the theory on a desired manifold and the $T \bar{T}$ operator is a composite of the stress tensor,
\begin{equation}
  T \bar{T} \equiv \det T_{\mu\nu} = \frac{1}{2} \epsilon^{\mu\rho} \epsilon^{\nu\sigma} T_{\mu\nu} T_{\rho\sigma}.
  \label{eqn:ttbar}
\end{equation}
The sign of $-\eta$ is the sign of the cosmological constant $\Lambda_{3}$ in the bulk dual (to be distinguished from the 2d cosmological constant $\Lambda_{2}$, a shorthand for the second term in the above differential equation).
$\eta = 0$ is the case of $\Lambda_{3} = 0$.

This deformation is only rigorously defined for all QFTs on flat two-dimensional manifolds \cite{Zamolodchikov:2004ce,Cavaglia:2016oda,Cardy:2018sdv,GST,LLST}, though attempts have been made to understand it in curved spaces \cite{Tolley:2019nmm,Mazenc:2019cfg, Caputa:2020lpa} and higher dimensions \cite{Taylor:2018xcy, Hartman:2018tkw, Belin:2020oib} in perturbation theory.
Here, we work only with holographic CFTs, where there is a large-$c$ semi-classical limit due to which there is no subtlety regarding well-definedness.
More specifically, we restrict to the case when the low-energy bulk effective theory is pure Einstein gravity {and the effects of matter fields (which contribute at order $c^0$ in the semi-classical expansion) are neglected}. There is no top-down example of such a system, and any explicit stringy construction is expected to contain a matter sector containing e.g. scalar moduli fields, but this is a useful starting point.

A differential equation is not sufficient to define a theory; we also need a boundary condition.
For $\eta = 1$, when the second term vanishes, there is an obvious boundary condition at $\lambda = 0$,
\begin{equation}
  Z_{\eta=1,\lambda} \xrightarrow{\lambda \to 0} Z_{CFT}.
  \label{eqn:ads-def-bd-cond}
\end{equation}
For $\eta \neq 1$, however, the $\Lambda_{2}$ term prevents us from taking this limit.
Since the $\Lambda_{2}$ term disappears at $\lambda \to \infty$, we may set a boundary condition in this limit instead,
\begin{equation}
  \lim_{\lambda \to \infty} Z_{\eta \neq 1, \lambda} = \lim_{\lambda \to \infty} Z_{\eta = 1, \lambda}
  \label{eqn:other-def-bd-cond}
\end{equation}
It was argued in \cite{Coleman:2021nor} that this boundary condition didn't result in a modular-invariant theory, and thus one has to add its modular images; we deal with this in section \ref{sec:saddle}.

The deformed path integral has the scale-invariance
\begin{equation}
  Z [\lambda,g_{2}] = Z \left[ e^{2 \Omega} \lambda, e^{2 \Omega} g_{2} \right],
  \label{eqn:scale-invariance}
\end{equation}
where $\Omega$ is a constant.
We will be interested in the case when the manifold is a torus or a cylinder.
Referring to the length of the spatial cycle as $2\pi R$, we account for the scale-invariance by defining the dimensionless quantity
\begin{equation}
    \hat{\lambda} \equiv \frac{\lambda}{R^2}.
\end{equation}
\cite{Coleman:2021nor} used a different dimensionless variable $y = \hat{\lambda}/(2\pi)^2$.

In the deformed CFT, the energies of the zero-momentum energy eigenstates on a circle of circumference  $2\pi R$ satisfying the boundary conditions \eqref{eqn:ads-def-bd-cond} and \eqref{eqn:other-def-bd-cond} are \cite{LLST}\footnote{Our $\lambda$ is $\pi/2$ times the $\lambda$ of \cite{LLST}.}
\begin{equation}
  E_{n}^{(S^{1})} (R,\lambda) = \frac{2\pi R}{2\lambda} \left\{ 1 - \sqrt{\eta - 4\mathcal{T} \hat{\lambda} + 4 \hat{\lambda}^{2} J^{2}} \right\}, \qquad \mathcal{T} = \frac{h + \bar{h} - \frac{c}{12}}{2\pi},\ J = \frac{h - \bar{h}}{2\pi}.
  \label{eqn:S1-En}
\end{equation}
where $\eta = +1$ in $AdS$, $\eta = -1$ in $dS$ and $\eta = 0$ in flat space.
With $\eta=1$, turning off the deformation gives $E_{n} \to 2\pi \mathcal{T}/R$.

This deformed theory turns out to have a semi-classical limit at large $c$, with the scaling $\lambda \to 0, c \to \infty,$ with $\lambda c $ finite.
The dual theory is defined by the Euclidean action
\begin{equation}
  I = - \frac{1}{16\pi G_N} \int_{\mathcal{M}_{3}} \left( R + \frac{2}{\ell^{2}} \eta \right) - \frac{1}{8\pi G_N} \int_{\partial \mathcal{M}_{3}} \left( K - \frac{1}{\ell} \right).
  \label{eqn:I-gen}
\end{equation}
Defining the quantity
\begin{equation}
  \kappa \equiv \frac{c}{24\pi} = \frac{\ell}{16\pi G_{N}},
  \label{eqn:kappa}
\end{equation}
the gravitational path integral dual to the 2d $T \bar{T} + \Lambda_{2}$-deformed path integral on a manifold $\mathcal{M}_{2}$ is defined by the above action and the Dirichlet boundary condition\footnote{In \cite{Kraus:2018xrn}, $4\kappa\lambda$ is set to $\ell^2$ and the flow of $\lambda$ is absorbed into $ds_{\mathcal{M}_2}^2$. In \cite{McGough:2016lol}, this equation is written in a coordinate-dependent form.}
\begin{equation}
  ds_{\mathcal{M}_{3}}^{2}\big|_{\partial \mathcal{M}_{3}} = \frac{\ell^{2}}{4\kappa\lambda} ds_{\mathcal{M}_{2}}^{2}.
  \label{eqn:dirichlet-bd-cond}
\end{equation}
For $\lambda > 0$, this is a cutoff at finite distance from bulk physics; $\lambda = 0, \eta=1$ is the usual asymptotic boundary of $AdS_{3}$.
$\lambda \to \infty$ is the case in which the length scale of the boundary goes to sub-AdS length scales so that the bulk `looks' flat regardless of cosmological constant.
This is why $\lambda \to \infty$ turns out to be a good boundary condition to define the $\eta \neq 1$ theory \cite{GST}.
We read off the boundary stress tensor from the bulk metric using the Brown-York formula along with the rescaling \eqref{eqn:dirichlet-bd-cond}
\begin{equation}
  T_{\mu\nu} = \frac{\ell}{2 \lambda} \left\{ K_{\mu\nu} - g_{\mu\nu} \left(K-\frac{1}{\ell}\right) \right\}.	
  \label{eqn:BY-formula}
\end{equation}

The eigenstates with energies \eqref{eqn:S1-En} correspond to some simple regions in the bulk.
Let us begin with the case of a spinless eigenstate in the $T \bar{T}$-deformed theory, i.e. with $\eta = 1$ and $J = 0$ and some arbitrary value of $\mathcal{T} > - \kappa$.
The boundary metric is
\begin{equation}
  ds^{2} = - dt^{2} + R^{2} d\theta^{2}, \quad \theta \sim \theta + 2\pi.
  \label{eqn:2d-g-cyl}
\end{equation}
To find the bulk dual, we embed this cylinder into a static $AdS_{3}$ spacetime with metric
\begin{equation}
  ds_{3}^{2} = - \frac{r^{2} - r_{h}^{2}}{\ell^{2}} dt_s^{2} + \frac{\ell^{2} dr^{2}}{r^{2} - r_{h}^{2}} + r^{2} d\phi^{2}, \qquad r_{h}^{2} > -\ell^2,\ \phi \sim \phi + 2\pi.
  \label{eqn:3d-g-ads}
\end{equation}
The case $r_{h}^{2} = -\ell^2$ is global $AdS$.
We can embed \eqref{eqn:2d-g-cyl} into \eqref{eqn:3d-g-ads} on a constant $r = r_{c}$ surface with the identifications
\begin{equation}
  r_c^2 = \frac{\ell^2}{4\kappa\hat{\lambda}}, \quad t = R \sqrt{1 - \frac{r_{h}^{2}}{r_{c}^{2}}}\ \frac{t_s}{\ell}, \quad \theta = \phi.
  \label{eqn:estate-embed}
\end{equation}
Finally, we have to impose for consistency that the energy given by the Brown-York formula matches the boundary energy.
The Brown-York formula gives for the energy
\begin{align}
  E = \int T^{t}_{t}  &= \frac{2\pi R}{2\lambda} \left\{ 1 - \sqrt{1 - 4 \kappa \hat{\lambda} \frac{r_{h}^{2}}{\ell^{2}}} \right\} \nonumber\\
  \Rightarrow \qquad \mathcal{T} &= \kappa \frac{r_{h}^{2}}{\ell^{2}} \qquad \text{or} \qquad h = \bar{h} = \frac{c}{24} \left( 1 + \frac{r_{h}^{2}}{\ell^{2}} \right).
  \label{eqn:t-rh-reln}
\end{align}
We can see that global $AdS_{3}$ is dual to the vacuum and that the Hawking-Page level $\mathcal{T} = \kappa$ is dual to the smallest canonically stable black hole with $r_{h} = \ell$ --- as expected.

\iffalse
From here on, we set $\ell = 1$ and $L = 2\pi$ except for the purpose of exhibiting some central results.
While it may seem odd to set two length scales to a number, remember that one is a length scale in the 2d theory and one is a length scale in the 3d theory.
To restore these length scales in the formulas below, we just need to plug in a power of either $l$ or $L/2\pi$ depending on whether we are looking at a 3d quantity or a 2d one (respectively).
\dg{Be more careful about flat space case.}
\fi

Similarly, we find that with $\eta = -1$, the real eigenstates $\mathcal{T} < - 1/4\hat{\lambda}$ are dual to a subregion of the static patch of $dS_{3}$ with a conical deficit,
\begin{equation}
  ds_{3}^{2} = - \left( 1 - \frac{r^{2}}{\ell^2} \right) dt_s^{2} + \frac{dr^{2}}{1-\frac{r^{2}}{\ell^2}} + r^{2} d\phi^2, \qquad r \in \left[ 0, r_{c} = \frac{\ell}{\sqrt{4\kappa\hat{\lambda}}} \right], \quad \phi \sim \phi + 2\pi \frac{(- \mathcal{T})}{\kappa}.
  \label{eqn:static-patch-old}
\end{equation}
Note that this patch never contains the cosmic horizon of $dS_{3}$; this is concomitant with the fact that there are no real states at and above the Hawking-Page level $\mathcal{T} \ge \kappa$, meaning that there are not enough real states to account for the entropy of the horizon. A description of the subregion containing the cosmic horizon in terms of dressed CFT microstates was given in \cite{Coleman:2021nor}, accounting for the entropy of the horizon.

It was argued in \cite{Donnelly:2018bef,LLST,Murdia:2019fax}, using replica trick arguments, that this duality also exhibits a version of the HRT formula, which states that the entanglement entropy of a region $A$ of the boundary theory is given by $1/4 G_N$ times the area of the minimal extermal surface anchored on $\partial A$.
It should be noted that none of these references proved that the $n$-replicated partition function has a well-defined bulk dual except in the case with a $U(1)$ symmetry; but we expect, due to arguments in \cite{LLST}, that there is a solution in the limit $n \to 1$ and so we should be able to derive an entanglement entropy (though not necessarily Renyi entropies).
Thus, we will use the HRT formula throughout this work.

It was shown in \cite{Dubovsky:2017cnj,Dubovsky:2018bmo} that the $T \bar{T}$ spectrum could be recovered from a topological gravity partition function.
In this partition function, we call the vielbeins corresponding to the `target space' (TS) torus on which the deformed theory lives $f^{a} = f_{\alpha}^{a} dX^{\alpha}$ and integrate over the vielbein $e^{a} = e_{\mu}^{a} dx^{\mu}$ of a `base space' (BS) torus and a map between the TS coordinates $X^{\alpha}$ and the BS coordinates $x^{\mu}$:
\begin{equation}
  Z_{\lambda} [f] = \int \frac{De DY}{\text{vol(diff)}}\ e^{- \frac{1}{2\lambda} \int \varepsilon_{ab} \left( dX - e \right)^{a} \wedge \left( dX - e \right)^{b}} Z_{0} [e], \qquad X^{a} \equiv f_{\alpha}^{a} \left( x^{\alpha} + Y^{\alpha} \right).
  \label{eqn:dghc}
\end{equation}
We can use the $Y$ integral and the diffeomorphism-invariance to localise $e$ on to a flat vielbein following \cite{Dubovsky:2018bmo}.
Alternatively, we can use the diffeomorphism-invariance to set $Y = 0$, obtaining the Freidel kernel \cite{Freidel:2008sh}.
We will use this latter gauge, and use the $c \to \infty, \lambda \to 0, c \lambda$ classical limit where the equation of motion is \cite{Mazenc:2019cfg}
\begin{align}
  e_{\mu}^{a} = f_{\mu}^{a} + \lambda (\det e) \varepsilon_{\mu\nu} \varepsilon^{ab} \langle T^{\nu}_{b} \rangle_{{}_{\text{\tiny CFT}}} (e),
  \label{eqn:kernel-eom}
\end{align}

\section{Bulk duals of the Thermofield Double state in the deformed theories} \label{sec:saddle}
In this section, we find the bulk dual of the thermofield double for all three values of $\eta$.
We define the bulk dual by finding the dominant bulk saddle contributing to the $T^2$ partition function, and using the Euclidean time-reflection-symmetric slice as initial data for real-time evolution.
In classic AdS/CFT this procedure gives two phases --- two entangled copies of thermal AdS at low temperatures and the two-sided eternal black hole at high temperatures.
We find such a Hawking-Page transition for all three cases of interest.

In this section, we take the Euclidean boundary to be $T^2$ with spatial cycle of circumference $2\pi R$ and time cycle of circumference $\beta$.
We take the non-rotating case for simplicity.
It will simplify matters to express our results in terms of the dimensionless quantities
\begin{align}
  \hat{\beta} \equiv \frac{\beta}{R}, \quad \hat{\lambda} \equiv \frac{\lambda}{R^{2}}, \quad \kappa \equiv \frac{c}{24\pi} = \frac{\ell}{16 \pi G_{N}}.
  \label{eqn:dimless-qtts}
\end{align}
In standard CFT notation the modular parameter of the torus is $\tau = i \hat{\beta}/2\pi$; we will use the $\hat{\beta}$ notation for consistency throughout, and stay away from the $\tau$ notation.

The CFT partition function in the large $c$ limit, and sufficiently far from $\beta = 2\pi R$, can be approximated as \cite{Hartman:2013qma}
\begin{equation}
    - \log Z = \min \left[ \beta E_{gs} (2\pi R), 2\pi R E_{gs} (\beta) \right],
    \label{eqn:S-Z-gen}
\end{equation}
where $E_{gs} (L)$ is the ground state energy on an $S^1$ of circumference $L$.
We will find that the bulk picture is consistent with this form for the partition function even in the deformed theories.
In the case of the CFT, $E_{gs} (L) = - (2\pi)^2 \kappa/L$, and we have
\begin{equation}
  \log Z_{CFT} = (2\pi)^2 \kappa \max \left( \frac{\hat{\beta}}{2\pi}, \frac{2\pi }{\hat{\beta}} \right).
    \label{eqn:cft-Z}
\end{equation}
The transition between these two phases happens at $\beta = 2\pi R$ in the CFT.

The two phases can be understood from the bulk point of view as the thermal AdS and BTZ phases respectively, and the transition between these is just the Hawking-Page transition.
The first phase, in which the ground state is propagating in the direct channel --- i.e. along the cycle that we have arbitrarily decided to call time ---, is dual to thermal AdS.
It has bulk Euclidean metric
\begin{equation}
  ds_3^2 = \left( 1 + \frac{\rho^2}{\ell^2} \right) dt_{gl,E}^2 + \frac{d\rho^2}{1 + \frac{\rho^2}{\ell^2}} + \rho^2 d\theta^2, \qquad \theta \sim \theta + 2\pi, t_{gl} \sim t_{gl} + \ell \hat{\beta}, \rho \in \left[ 0, \frac{R}{\epsilon} \right].
    \label{eqn:gl-ads}
\end{equation}
The induced metric at the boundary is $ds_{2}^{2}/\epsilon^{2}$, where $ds_{2}^{2}$ is the $T^{2}$ metric; the $\rho$ coordinate of the boundary and the periodicity of $t_{gl}$ are fixed by this requirement.
Cutting this at the reflection-symmetric surface $t_{gl,E} \in \left\{ 0, \frac{\ell}{2} \hat{\beta} \right\}$ and evolving in real time, we find that the Lorentzian geometry is two copies of global AdS.
The Euclidean bulk action for this geometry can be found and it agrees exactly with $\beta E_{gs} (2\pi R)$.

The second phase is the non-rotating BTZ black hole.
This corresponds to the vacuum propagating in the cross channel, i.e. along the cycle that we have arbitrarily defined to be space.
The bulk Euclidean metric is
\begin{equation}
    ds_3^2 = \frac{r^2 - r_h^2}{\ell^2} dt_{s,E}^2 + \frac{\ell^2 dr^2}{r^2 - r_h^2} + r^2 d\phi^2, \qquad \phi \sim \phi + 2\pi, r \in \left[ 0, \frac{R}{\epsilon} \right], t_{s,E} \sim t_{s,E} + \frac{2\pi \ell^2}{r_h}.
    \label{eqn:btz}
\end{equation}
The temperature and horizon radius are related by
\begin{equation}
  \hat{\beta} = \frac{2\pi \ell}{r_h}.
    \label{eqn:btz-beta-rh}
\end{equation}
Evolving in real time from the reflection-symmetric slice $t_{s,E} \in \left\{ 0, \frac{\pi \ell^2}{r_h} \right\}$ results in the two-sided eternal black hole geometry.
For future reference, we also define the `Penrose diagram' coordinates,
\begin{align}
  \frac{s \pm y}{2} \equiv \tan^{-1} w^{\pm}, \qquad &w^{\pm} \equiv \pm \sqrt{\frac{r - r_{h}}{r + r_{h}}} e^{\pm \frac{r_{h} t_{s}}{\ell^{2}}}, \nonumber\\
  ds_{3}^{2} &= \frac{\ell^{2} \left( - ds^{2} + dy^{2} \right) + r_{h}^{2} \cos^{2} s d\phi^{2}}{\cos^{2} y}, \qquad s,y \in \left[ - \frac{\pi}{2}, \frac{\pi}{2} \right].
  \label{eqn:penrose-coords}
\end{align}

The bulk action can be found to be $2\pi R E_{gs}(\beta)$, consistent with the saddle being nothing but the ground state propagating in the cross-channel; this can also be seen geometrically using the coordinate transformations
\begin{equation}
    \tilde{\theta} = \frac{r_h}{\ell^2} t_{s,E}, \quad \tilde{\rho} = \sqrt{\frac{r^2}{r_h^2} - 1} \quad \tilde{t}_{gl,E} = \frac{r_h}{\ell} \phi,
    \label{eqn:s-trans-coord}
\end{equation}
which brings \eqref{eqn:btz} to the form \eqref{eqn:gl-ads} with $\beta \leftrightarrow 2\pi R$.
The principle to derive this coordinate transformation is (a) flip space and time and (b) calling the new spatial coordinate $\phi$ and demanding that it have periodicity $2\pi$, define the new radial coordinate to be the square root of $g_{\phi\phi}$.

\subsection{Finite-cutoff AdS} \label{ssec:ads-saddle}
Now we calculate the bulk duals of the thermofield double in the finite-cutoff AdS theory.
To do this, we have to calculate the actions of the finite-cutoff thermal AdS and BTZ saddles as a function of boundary temperature and the transition between them.

We start with the thermal AdS saddle.
Embedding a $T^{2}$ with cycles $2\pi R$ and $\beta$ into the thermal AdS metric \eqref{eqn:gl-ads} with the rule \eqref{eqn:dirichlet-bd-cond} gives the coordinate ranges
\begin{equation}
  \rho \le \rho_{c} = \frac{\ell}{\sqrt{4 \kappa \hat{\lambda}}}, \quad t_{gl,E} \sim t_{gl,E} + \frac{\ell}{\sqrt{1 + 4\kappa \hat{\lambda}}} \hat{\beta}.
  \label{eqn:th-ads-fin-cut}
\end{equation}
This reduces to the ranges in \eqref{eqn:gl-ads} when $\lambda = \frac{\ell^{2} \epsilon^{2}}{4\kappa} \ll \ell^{2}$.
The on-shell action is
\begin{equation}
  I_{gl} \left[ \hat{\beta}, \hat{\lambda} \right] = \pi \frac{\hat{\beta}}{\hat{\lambda}} \left\{ 1 - \sqrt{1 + 4 \kappa \hat{\lambda}} \right\} \quad \xrightarrow{\hat{\lambda} \to 0} - 2 \pi \kappa \hat{\beta}
  \label{eqn:th-ads-I}
\end{equation}

As for the BTZ saddle, the same rules result in the coordinate ranges
\begin{align}
  r \le r_{c} \equiv \frac{\ell}{\sqrt{4 \kappa \hat{\lambda}}}, \quad t_{s,E} \sim t_{s,E} + \frac{2\pi \ell^{2}}{r_{h}}, \quad r_{h} = \frac{\ell}{\sqrt{(\hat{\beta}/2\pi)^{2} + 4 \kappa \hat{\lambda}}}, \quad \hat{\beta} = \frac{2\pi \ell}{r_{h}} \sqrt{1 - \frac{r_{h}^{2}}{r_{c}^{2}}}.
  \label{eqn:btz-fin-cut}
\end{align}
The relation between Schwarzschild time and boundary time is
\begin{equation}
  t = R \sqrt{1 - \frac{r_{h}^{2}}{r_{c}^{2}}} \frac{t_{s}}{\ell} = \frac{\beta/2\pi}{\sqrt{4 \kappa \hat{\lambda} + \left( \hat{\beta}/2\pi \right)^{2}}} \frac{t_{s}}{\ell}
    \label{eqn:t-ts-final}
\end{equation}
The on-shell action is
\begin{equation}
  I_{BTZ} \left[ \hat{\beta}, \hat{\lambda} \right] = \pi \frac{\hat{\beta}}{\hat{\lambda}} \left\{ 1 - \sqrt{1 + \frac{4\kappa \hat{\lambda}}{(\hat{\beta}/2\pi)^{2}}} \right\} \qquad \xrightarrow{\hat{\lambda} \to 0} - \left( 2\pi \right)^{2} \kappa \frac{2\pi}{\hat{\beta}}.
  \label{eqn:btz-I}
\end{equation}

The partition function from the two saddles satisfies
\begin{equation}
  \log Z = - \min \left( I_{gl}, I_{BTZ} \right) = \max \left[ \beta E_{gs} (\lambda, 2\pi R), 2\pi R E_{gs} (\lambda,\beta) \right].
  \label{eqn:Z-AdS}
\end{equation}
Here, $E_{gs}$ is the ground state energy of the deformed theory \eqref{eqn:S1-En} with $\eta=1$. The Hawking-Page transition continues to be at $\beta = 2\pi R$ for all $\lambda > 0$.

\subsection{Finite-cutoff dS} \label{ssec:ds-saddle}
Turning to $\eta = -1$, we find a Hawking-Page transition between two different patches of dS${}_{3}$, the `pole patch' and the `cosmic horizon patch' \cite{Coleman:2021nor}.
In \cite{Coleman:2021nor} we only considered patches of global dS${}_{3}$, since our interest was in understanding properties of 3d gravity.
In this work, we consider the bulk to be a calculational tool to understand properties of the 2d theory, and so we will consider a more general class of geometries.

We consider Euclidean dS${}_{3}$ in static patch coordinates,
\begin{equation}
  ds_{3}^{2} = \frac{\rho_{h}^{2} - \rho^{2}}{\ell^{2}} dt_{s,E}^{2} + \frac{\ell^{2} d\rho^{2}}{\rho_{h}^{2} - \rho^{2}} + \rho^{2} d\phi^{2}, \qquad \phi \sim \phi + 2\pi, t_{s,E} \sim t_{s,E} + \mathbb{b}.
  \label{eqn:dS-static}
\end{equation}
Global dS${}_{3}$ corresponds to $\rho_{h} = \ell$ and $\mathbb{b} = 2\pi \ell$; in \cite{Coleman:2021nor} we stuck to this case.

There are two different patches of dS${}_{3}$ with a $T^{2}$ boundary
\begin{align}
  \text{Pole patch:} \qquad \rho \in \left[ 0, \rho_{c} \right], \ \rho_{h} &= \ell, \ \mathbb{b} \text{ free}. \nonumber\\
  \text{Cosmic Horizon (CH) patch:} \qquad \rho \in \left[ \rho_{c}, 1  \right], \ \mathbb{b} &= \frac{2\pi \ell^{2}}{\rho_{h}}, \ \rho_{h} \text{ free}.
  \label{eqn:ds-patches}
\end{align}
Smoothness at the pole $\rho = 0$ requires $\rho_{h} = \ell$, and smoothness at the horizon $\rho = \rho_{h} $ requires $\mathbb{b} = 2\pi \ell^{2}/\rho_{h}$.
We only need to impose one of these two conditions in either case, and so one of $\rho_{h}, \mathbb{b}$ is free in either patch; they will be fixed below by the boundary temperature.

The pole patch is analogous to the thermal AdS saddle.
The relation between bulk and boundary parameters is
\begin{equation}
  \rho_{c} = \frac{\ell}{\sqrt{4\kappa \hat{\lambda}}}, \qquad \mathbb{b} = \frac{\ell}{\sqrt{4\kappa \hat{\lambda} - 1}} \hat{\beta}, \qquad t_{s} = \frac{\ell}{\sqrt{4\kappa \hat{\lambda} - 1}} \frac{t}{R}.
  \label{eqn:pole-bulk-bd-dict}
\end{equation}
The Euclidean gravitational action of this saddle is
\begin{align}
  I_{P} &= - \frac{1}{16\pi G_{N}} \int_{0}^{2\pi} d\phi \int_{0}^{\mathbb{b}} dt_{s,E} \int_{0}^{\rho_{c}} d\rho\ \left( R - \frac{2}{\ell^{2}} \right) - \frac{1}{8\pi G_{N}}\int_{0}^{2\pi} d\phi \int_{0}^{\mathbb{b}} dt_{s,E}\ \left( K - \frac{1}{\ell} \right) \nonumber\\[0.5em]
  &= \frac{2 \pi \hat{\beta}}{2 \hat{\lambda}} \left\{ 1 - \sqrt{4\kappa \hat{\lambda} - 1} \right\} \nonumber\\[0.5em]
  &= \beta E_{gs} \left( 2\pi R, \lambda \right).
  \label{eqn:pole-path-action}
\end{align}
Here, $E_{gs}$ is the deformed ground state energy \eqref{eqn:S1-En} with $\eta = -1$.

The cosmic horizon patch is analogous to the black hole saddle in AdS.
The relation between bulk and boundary parameters is
\begin{equation}
  \rho_{c} = \frac{\ell}{\sqrt{4\kappa \hat{\lambda}}}, \qquad \rho_{h} = \frac{\ell}{\sqrt{4\kappa \hat{\lambda} - (\hat{\beta}/2\pi)^{2}}}, \qquad t_{s} = \frac{t \ell}{R}\sqrt{\frac{4\kappa \hat{\lambda}}{\left( \hat{\beta}/2\pi \right)^{2}} - 1} .
  \label{eqn:ch-bulk-bd-dict}
\end{equation}
The case $\rho_{h} < \ell$ corresponds to a conical deficit and the case $\rho_{h} > \ell$ corresponds to a conical \emph{excess}.
A conical excess, by the Einstein equations, is sourced by negative energy and therefore this case is forbidden for global dS; however, it is allowed for us because the conical excess is at $\rho = 0$ and so is in the excluded region.

It will also be useful to note down Penrose diagram coordinates that cover the whole of Lorentzian dS${}_{3}$,
\begin{align}
  \frac{s \pm y}{2} \equiv \tan^{-1} w^{\pm}, \qquad &w^{\pm} \equiv \pm \sqrt{\frac{\rho_{h} - \rho}{\rho_{h} + \rho}} \, e^{\pm \frac{\rho_{h} t_{s}}{\ell^{2}}} \nonumber
\end{align}
\begin{align}
  ds_{3}^{2} &= \frac{\ell^{2} (- ds^{2} + dy^{2}) + \rho_{h}^{2} \cos^{2} y d\phi^{2}}{\cos^{2} s}.
  \label{eqn:ds-penrose}
\end{align}

One can easily generalise these considerations to add rotation.
The metric can be found in \cite{Bousso:2001mw}; with the difference that we can also allow negative masses for the same reason as above.
We will not explore this solution further here.

The Euclidean gravitational action is
\begin{align}
  I_{CH} &= \frac{2\pi \hat{\beta}}{2\hat{\lambda}} \left\{ 1 - \sqrt{\frac{4\kappa \hat{\lambda}}{\left( \hat{\beta}/2\pi \right)^{2}} - 1} \right\}.
  \label{eqn:ch-action}
\end{align}

The partition function from these two contributions satisfies, same as \eqref{eqn:Z-AdS},
\begin{equation}
  \log Z = \max \left[ \beta E_{gs} \left( \lambda, 2\pi R \right), 2\pi R E_{gs} \left( \lambda, \beta \right) \right].
  \label{eqn:Z-dS}
\end{equation}
Yet again, we find that the Hawking-Page transition is at $\beta = 2\pi R$.

\subsection{Flat space} \label{ssec:flat-saddle}
In the case of a flat 3d bulk, which corresponds to $\eta = 0$ on the boundary, it is harder to make sense of the Dirichlet boundary condition \eqref{eqn:dirichlet-bd-cond} since there is no bulk length scale.
To make sense of this, we take the flat limit as
\begin{equation}
  \ell \to \infty, c \lambda \to \infty, \qquad \tilde{\lambda} = \frac{4\kappa\lambda}{\ell^{2}} > 0.
  \label{eqn:flat-limit}
\end{equation}
The Dirichlet boundary condition becomes
\begin{equation}
  ds_{\mathcal{M}_{3}}^{2} \Big|_{\partial \mathcal{M}_{3}} = \frac{1}{\tilde{\lambda}} ds_{\mathcal{M}_{2}}^{2}.
  \label{eqn:flat-dirichlet-bd-cond}
\end{equation}

There are two ways to embed a torus into flat space of dimensions $2\pi R \times \beta$ into flat space.
The first way, which we'll call the ``thermal flat space'' case, has bulk metric
\begin{equation}
  ds_{3}^{2} = dt_{m,E}^{2} + dr^{2} + r^{2} d\phi^{2}, \qquad \phi \sim \phi + 2\pi,\ r \in \left[ 0, \frac{R}{\sqrt{\tilde{\lambda}}} \right],\ t_{m,E} \sim t_{m,E} + \frac{\beta}{\sqrt{\tilde{\lambda}}}.
  \label{eqn:thermal-flat}
\end{equation}
The relation between bulk and boundary times in this case is
\begin{equation}
  t_{m} = t \sqrt{\tilde{\lambda}} .
  \label{eqn:thermal-fl-ts}
\end{equation}
The bulk action is 
\begin{align}
  I_{th-fl} &= - \frac{1}{16\pi G_{N}} \int_{0}^{2\pi} d\phi \int_{0}^{\beta/\sqrt{\tilde{\lambda}}} dt_{m,E} \int_{0}^{R/\sqrt{\tilde{\lambda}}} dr\ \sqrt{g} \, R - \frac{1}{8 \pi G_{N}} \int_{0}^{2\pi} d\phi \int_{0}^{\beta/\sqrt{\tilde{\lambda}}} dt_{m,E} \sqrt{h} \, K \nonumber\\[1em]
  &= - \frac{\beta}{4 G_{N} \sqrt{\tilde{\lambda}}}.
  \label{eqn:th-fl-I}
\end{align}

Same as in the case with a non-zero cosmological constant, the other embedding can be found by exchanging time and space.
We find that the resulting space is a patch of Euclidean Rindler space,
\begin{equation}
  ds^{2} = r^{2} dt_{r,E}^{2} + dr^{2} + \frac{R^{2}}{\tilde{\lambda}} d\phi^{2}, \qquad \phi \sim \phi + 2\pi, r \in \left[ 0, \frac{\beta/2\pi}{\sqrt{\tilde{\lambda}}} \right], t_{r,E} \sim t_{r,E} + 2\pi.
  \label{eqn:rindler}
\end{equation}
The relation between bulk and boundary times is
\begin{equation}
  t_{r} = \frac{t}{\beta/2\pi}.
  \label{eqn:rindler-ts}
\end{equation}
The bulk action is
\begin{equation}
  I_{R} = - \frac{2\pi R}{4 G_{N} \sqrt{\tilde{\lambda}}}.
  \label{eqn:R-I}
\end{equation}
Yet again, we find that the Hawking-Page transition between the two phases is at $\beta = 2\pi R$.

Typically, compactifying the transverse direction in Rindler space is not stable to gravitational fluctuations, since a change in the compactification scale is a massless direction in configuration space, see \cite{Goheer:2003tx}.
However, with our finite-radius boundary condition there is no such problem.

With the actions computed, we can begin to understand the entanglement spectrum of the entanglement between the two boundaries. The Renyi entropy in the high-temperature Rindler Horizon (RH) phase is given by
\begin{equation}
  S_{n} = \frac{\log Z(n \beta) - n \log Z(\beta)}{1-n} = \frac{2\pi R}{4 G_{N} \sqrt{\tilde{\lambda}}} = \frac{A_{RH}}{4G_{N}}.
  \label{eqn:flat-renyi}
\end{equation}
Thus, we see that it is completely independent of $n$.
An inverse Laplace transform explains why this is the case: the $\beta$-independence of the partition function means that the density of states is
\begin{equation}
  \rho(E) = \int_{-i\infty}^{i\infty} d\beta\ e^{\beta E} Z(\beta) = \delta(E) e^{\frac{A_{RH}}{4 G_{N}}}.
  \label{eqn:rindler-dos}
\end{equation}
In other words, the only contributing state is the vacuum -- one cannot excite the system to a higher mass, presumably because there are no $M>0$ Schwarzschild black holes in $(2+1)$ dimensions, nor conical deficit states.

\section{Entanglement Tsunami in $AdS_3$} \label{sec:ads}

We turn now to studying the spread of entanglement in cutoff $AdS_3$. 
At fixed $\hat{\lambda}$ (fixed cutoff radius $r_c$), we construct the thermofield double state of a given inverse temperature $\beta$ and evolve forward in time.

\subsection{One Interval on Each Side}
\begin{figure}[ht!]
    \centering
    \includegraphics[width=.35\textwidth]{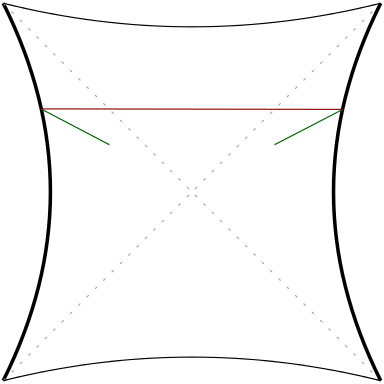}
    \hspace{.05\textwidth}
    \includegraphics[width=.45\textwidth]{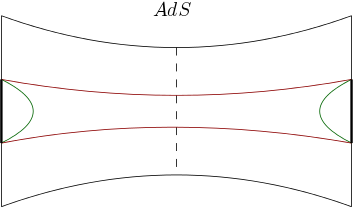}
    \caption{The two types of RT surfaces in AdS. Left: On the Penrose diagram. Right: View from the ``top'' of the Penrose diagram.}
    \label{fig:ads-rt}
\end{figure}

We consider two intervals of size $\phi_0$, each one centered around $\phi=0$ on either the left or right boundary.
We compute the EE as a function of Lorentzian time $t$, using the HRT formula.
For discussions of the HRT formula in the deformed theory see \cite{Donnelly:2018bef,Murdia:2019fax}.
There are in general two topologies for the RT surface: a factorised one that connects the two ends of each interval, and an entangled one that connects the left interval to the right interval. The factorised extremal surface is symmetric under Schwarzschild time translations and lives entirely within the $r$-$\phi$ plane. By contrast, the entangled surface is symmetric under angular rotations and lives on the $r$-$t$ plane.
See figure \ref{fig:ads-rt}.

Let us first calculate the area of the factorised surface.
Since the metric is independent of $t$, the surface sits at fixed $t$ and the extremal-area condition becomes the extremisation of the area functional
\begin{align}
    A_{fact} = 2 \int d\xi \sqrt{g_{\mu\nu}\dfrac{dx^\mu}{d\xi}\dfrac{dx^\nu}{d\xi}} = \int d\phi \sqrt{\frac{\ell^2}{r^2 - r_h^2} r'(\phi)^2 + r^2} \equiv \int d\phi \, L.
    \label{eqn:A-fact}
\end{align}
The factor of two comes from the fact that there are two copies of the surface, one in each exterior.
Because of the $\phi$ translation symmetry,  there is an integral of motion given by
\begin{equation}
   c = r' \dfrac{\partial L}{\partial r'} - L = \frac{r^2}{\sqrt{r^2 + \frac{\ell^2 r'(\phi)^2}{r^2 - r_h^2}}} \qquad \Rightarrow \qquad r' = \pm \frac{r}{\ell} \sqrt{(r^2 - r_h^2) \left( \frac{r^2}{c^2} - 1 \right)}.
   \label{eqn:fact-soln}
\end{equation}
The surface has $r' = 0$ at the turning point $r = c \equiv r_0$, and the sign of $r'$ is different on the two different sides of the turning point.

The full trajectory is
\begin{equation}
  \phi (r) = \frac{\ell}{r_{h}} \tanh^{-1} \left( \frac{r_{h}}{r_{0}} \sqrt{\frac{r^{2} - r_{0}^{2}}{r^{2} - r_{h}^{2}}} \right)
  \label{eqn:ads-rt-conn}
\end{equation}
The turning point $r = r_{0}$ is determined by the width of the interval $\phi_{0}$, by the relation $\phi(r_{c}) = \phi_{0}/2$.
The solution is
\begin{equation}
  r_{0} (\phi_{0}) = \frac{r_{c} r_{h}}{\sqrt{r_{h}^{2} + \left( r_{c}^{2} - r_{h}^{2} \right) \tanh^{2} \left( \frac{r_{h} \phi_{0}}{2\ell} \right) }} \quad \xrightarrow{r_c \to \infty} r_{h} \coth \frac{r_h \phi_0}{2\ell} = \frac{2\pi \ell}{\hat{\beta}} \coth \frac{\phi_0/2}{\hat{\beta}/2\pi}.
  \label{eqn:ads-turnaround}
\end{equation}
Integrating \eqref{eqn:A-fact} with this solution, we find
\begin{align}
  S_{fact} = \frac{A_{fact}}{4 G_N} &= \frac{c}{3} \log \left\{\frac{\sqrt{r_{c}^{2} - r_{h}^{2}} + \sqrt{r_{c}^{2} - r_{0}^{2}}}{\sqrt{r_{0}^{2} - r_{h}^{2}}} \right\}^2 \nonumber\\[1em]
  &= \frac{2 c}{3} \sinh^{-1} \left\{ \frac{\hat{\beta}/2\pi}{\sqrt{4\kappa \hat{\lambda}}} \sinh \frac{R\phi_{0}/2}{\sqrt{(\beta/2\pi)^{2} + 4\kappa \lambda}} \right\} \nonumber\\[1em]
  & \qquad \xrightarrow{4 \kappa \lambda = \epsilon^{2} \ll 1} \frac{2c}{3} \log \left\{ 2 \frac{\beta}{2\pi \epsilon} \sinh \frac{R\phi_0/2}{\beta/2\pi} \right\}.
  \label{eqn:ads-conn-A}
\end{align}

Now we turn to the connected extremal surface.
It has two components, each extending from one endpoint on the left boundary to the corresponding endpoint on the right boundary.
To solve for its position, we go to maximally extended coordinates; in this case we use the Penrose diagram coordinates $s,y$ \eqref{eqn:penrose-coords}.
It is easily seen that the surface $s = s_0, \phi = \phi_0$ is extremal, since the induced metric only depends on $y$.
Two copies of this geodesic at $\phi = \pm \phi_0/2$ gives the connected HRT surface of interest.

The end-points of the geodesic in these coordinates are given by
\begin{align}
  \tan \frac{s_0 \pm y_0}{2} &= \pm \sqrt{\frac{r_c - r_h}{r_c + r_h}} \, e^{\pm \frac{r_h t_s}{\ell^2}}, \qquad s_{0} = \sqrt{1 - \frac{r_{h}^{2}}{r_{c}^{2}}}\, \sinh{\left(\frac{r_{h} t_{s}}{\ell^{2}}\right)},\ y_{0} = \sqrt{\frac{r_{c}^{2}}{r_{h}^{2}} - 1} \,\cosh{\left(\frac{r_{h} t_{s}}{\ell^{2}}\right)}.
\end{align}
Integrating the area $A_{conn} = 2 \int_{-y_{0}}^{y_{0}} \sec y \, dy$, we find
\begin{align}
  A_{conn} %= 2 \int_{0}^{y_{c}} \sqrt{g_{yy}} dy 
  &= 4 \sinh^{-1} \left[ \sqrt{\frac{r_{c}^{2}}{r_{h}^{2}} - 1} \cosh \left(\dfrac{r_{h} t_s}{\ell^2}\right) \right] \nonumber\\[1em]
  \Rightarrow S_{conn} &= \frac{2 c}{3} \sinh^{-1} \left[\sqrt{\frac{\left( \hat{\beta}/2\pi \right)^{2}}{4 \kappa \hat{\lambda}} - 1} \,\, \cosh{\left(\frac{t}{\beta/2\pi}\right)}  \right].
  \label{eqn:ads-disc-A}
\end{align}

We thus find for the entropy,
\begin{align}
  S_{E} = 2 \frac{c}{3} \sinh^{-1} \min \left\{ \frac{\hat{\beta}/2\pi}{\sqrt{4\kappa \hat{\lambda}}} \sinh \frac{R \phi_{0}/2}{\sqrt{\left( \beta/2\pi \right)^{2} + 4\kappa\lambda} } , \sqrt{\frac{\left( \hat{\beta}/2\pi \right)^{2}}{4\kappa \hat{\lambda}} - 1} \cosh \frac{t}{\beta/2\pi} \right\}.
  \label{eqn:one-int-final}
\end{align}
The locus $\phi_{*} (t)$ of the transition between these two phases is
\begin{align}
  \sinh^{2} \frac{R \phi_{*}/2}{\sqrt{\left( \beta/2\pi \right)^{2} + 4\kappa \lambda}} = \left[ 1 - \frac{4\kappa \hat{\lambda}}{(\hat{\beta}/2\pi)^{2}} \right] &\cosh^{2} \left(\frac{t}{\beta/2\pi}\right) \nonumber\\[0.5em]
  \xrightarrow{R\phi_{*},t \,\gg\,\frac{\beta}{2\pi}, \sqrt{4\kappa \lambda}} \qquad \frac{R\phi_{*}}{2} &\approx \sqrt{1 + \frac{4\kappa \hat{\lambda}}{\left( \hat{\beta}/2\pi \right)^{2}}}\ \left[ t - \frac{\beta}{2\pi} \log \frac{1}{\sqrt{1 - \frac{4\kappa \hat{\lambda}}{\left( \hat{\beta}/2\pi \right)^{2}}}} \right].
  \label{eqn:qp-traj-1}
\end{align}
The slope of this locus in the $(\phi,t)$ plane at late enough times is
\begin{equation}
  \frac{d R \phi_{*}/2}{d t} = \sqrt{1 + \frac{(\hat{\beta}/2\pi )^2}{4\kappa \hat{\lambda}}} = \frac{1}{\sqrt{1 - \frac{r_{h}^{2}}{r_{c}^{2}}}} > 1.
  \label{eqn:v-def-qp}
\end{equation}
It should be noted that the slope in bulk coordinates $\frac{dR \phi_{*}/2}{dt_{s}}$ is exactly 1.
The trajectory of the excitations found in \cite{McGough:2016lol} also had the same slope.

We can account for this by an entanglement tsunami picture as follows.
Imagine that the tsunami emanates from the end-points of the interval following the trajectory (note the similarity to \eqref{eqn:qp-traj-1})
\begin{align}
  E_{\pm} (t) &= \left( \pm \frac{\phi_{0}}{2} - \Delta \phi(t)\, ,\, \pm \frac{\phi_{0}}{2} + \Delta \phi(t) \right) \nonumber\\
  &\Delta \phi (t) \equiv \sqrt{\left( \hat{\beta}/2\pi \right)^{2} + 4\kappa \hat{\lambda}}\, \sinh^{-1} \left\{ \sqrt{1 - \frac{4\kappa \hat{\lambda}}{\left( \hat{\beta}/2\pi \right)^{2}}} \cosh \frac{t}{\beta/2\pi} \right\}.
  \label{eqn:tsunami-traj}
\end{align}
Then, the entanglement is
\begin{align}
  S_{E} = \frac{2c}{3} \sinh^{-1} \left\{ \frac{\hat{\beta}/2\pi}{\sqrt{4\kappa \hat{\lambda}}} \sinh \frac{\frac{1}{4} \mathrm{vol} \left( E(t) \cap A \right)}{\sqrt{\left( \beta/2\pi \right)^{2} + 4\kappa\lambda}} \right\}.
  \label{eqn:tsunami-ent-1-int}
\end{align}
This follows from the fact that $\mathrm{vol} \left( E(t) \cup A \right) = 4 \Delta\phi$; the factor of $4$ arises from the four end-points of $A$.
This is strictly true only for the case when $A$ covers less than half the spatial slice on each side; more generally, we should make the replacement $\mathrm{vol} (E(t) \cap A) \to \min \left[ \mathrm{vol} \left( E(t) \cap A \right), \mathrm{vol} \left( E(t) \cap A^{c} \right) \right]$.

\begin{figure}[h!]
  \centering
  \begin{tikzpicture}
    \draw[->] (0,0) -- node[below right,pos=1] {$\phi$} (4,0);
    \draw[thick] (1.5,0) -- node[below] {$A$} (2.5,0);
    \draw[dashed] (1.5,0) -- (1.5,3);
    \draw[dashed] (2.5,0) -- (2.5,3);
    \draw[->] (-1,0) -- node[left] {$t$} (-1,3);
    \fill[brown,opacity=.2] (1.5,0) to[out=80,in=-150] (2,1) to[out=-30,in=100] (2.5,0) to[out=80,in=-150] (4,2) -- (4,3) -- (0,3) -- (0,1.5) to[out=-30,in=100] (1.5,0);
  \end{tikzpicture}
  \caption{The entanglement tsunami that reproduces the entropy \eqref{eqn:one-int-final}.}
  \label{fig:one-int-tsunami}
\end{figure}
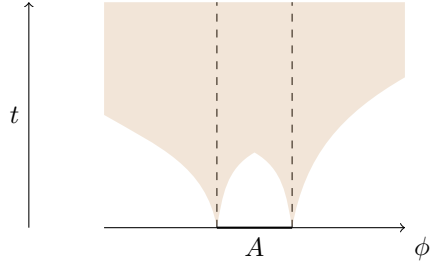

Of course, the tsunami ansatz for the entanglement of one interval on each side is a case where the number of parameters equals the number of data points.
As long as the entanglement entropy takes the form
\begin{equation}
  S_{E} = \min \left( f(\phi_{0}/2), g(t) \right),
  \label{eqn:gen-form}
\end{equation}
as it is bound to in a holographic set-up, we can define the entanglement tsunami in a manner similar to \eqref{eqn:tsunami-traj} with
\begin{align}
  \Delta \phi = f^{-1} \circ g(t), \qquad S_{E} = f \left( \frac{1}{4} \mathrm{vol} \left( E(t)  \cap A \right) \right).
  \label{eqn:tsunami-form-gen}
\end{align}
To see if this form has any meaning, we need to consider the case of multiple intervals.
We now turn to the case of two intervals to see whether the ansatz truly holds up.

\subsection{Two Intervals on Each Side} \label{ssec:two-int-ads}
We now consider the case when $A = A_{1,l} \cup A_{2,l} \cup A_{1,r} \cup A_{2,r}$ , a union of two intervals on each side.
We take $A_{1,l} = \left( - \delta\phi/2 - \phi_{1}, -\delta\phi/2 \right)$ and $A_{2,l} = \left( \delta\phi/2, \delta\phi/2 + \phi_{2} \right)$, and take $A_{1,r}, A_{2,r}$ to be the symmetrically placed intervals on the right system.
Without loss of generality, we take
\begin{equation}
  \phi_{1} \le \phi_{2} \le \pi, \qquad 2\pi - \left( \delta\phi + \phi_{1} + \phi_{2} \right) \ge \delta \phi.
  \label{eqn:two-int-conds}
\end{equation}
The first inequality says that $A_{1,l}$ is the smaller of the two intervals; and the second that $\delta\phi$ is the smaller of the angular gaps between the two intervals.

\begin{figure}[h!]
  \centering
  \includegraphics[width=.7\textwidth]{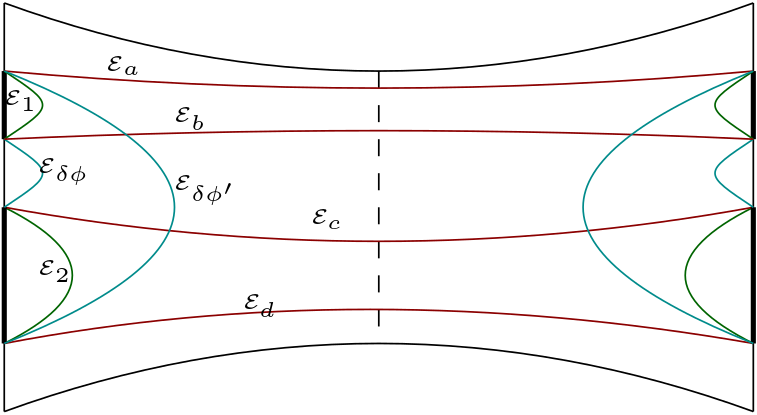}
  \caption{The possible geodesics in the two interval case.}
  \label{fig:ads-2-int}
\end{figure}

There are five possible extremal surfaces, each a combination of the types of surfaces we considered in the case of one interval on each side.
The various geodesics are shown in figure \ref{fig:ads-2-int}.
The candidate extermal surfaces are
\begin{align}
  X_{1} &= \mathcal{E}_{a} \cup \mathcal{E}_{b} \cup \mathcal{E}_{c} \cup \mathcal{E}_{d} \nonumber\\
  X_{2} &= \mathcal{E}_{1,l} \cup \mathcal{E}_{1,r} \cup \mathcal{E}_{c} \cup \mathcal{E}_{d} \nonumber\\
  X_{3} &= \mathcal{E}_{\delta\phi,l} \cup \mathcal{E}_{\delta\phi,r} \cup \mathcal{E}_{a} \cup \mathcal{E}_{d} \nonumber\\
  X_{4} &= \mathcal{E}_{1,l} \cup \mathcal{E}_{1,r} \cup \mathcal{E}_{2,l} \cup \mathcal{E}_{2,r} \nonumber\\
  X_{5} &= \mathcal{E}_{\delta\phi,l} \cup \mathcal{E}_{\delta\phi,r} \cup \mathcal{E}_{\delta\phi',l} \cup \mathcal{E}_{\delta\phi',r}.
  \label{eqn:ads-2-int-rts}
\end{align}
The areas of $\mathcal{E}_{a,b,c,d}$ are given by \eqref{eqn:ads-disc-A}.
The areas of $\mathcal{E}_{1,2,\delta\phi,\delta\phi'}$ are given by \eqref{eqn:ads-conn-A} with $\phi_{0}$ replaced by the angular width indicated by the subscript; we define $\delta\phi' = \min \left( 2\pi - \delta\phi-\phi_{1}-\phi_{2}, \delta\phi+\phi_{1}+\phi_{2}\right)$.
There are a few more possible combinations of these geodesics that are candidate extermal surfaces, but they can never be minimal given the assumptions \eqref{eqn:two-int-conds}.

Supposing $\phi_{1} + \phi_{2} < \pi$, the dominant HRT surfaces are $X_{1} \to X_{2} \to X_{4}$ as time progresses.
In the other case, where $\phi_{1} + \phi_{2} > \pi$, the dominant HRT surfaces are $X_{1} \to X_{3} \to X_{5}$ as time progresses.
Here, all we have used is that \eqref{eqn:ads-conn-A} is a monotonically increasing function of the angular width.
The reason for the impossibility of a transition between $X_2,X_3$ is that $\mathrm{Area} (X_{2} ) - \mathrm{Area} (X_{3}) = 2\mathrm{Area} \left( \mathcal{E}_{1} \right) - 2 \mathrm{Area} \left( \mathcal{E}_{\delta\phi} \right)$ is time-independent and so time-evolution cannot change its sign; similarly with $X_{4}, X_{5}$.

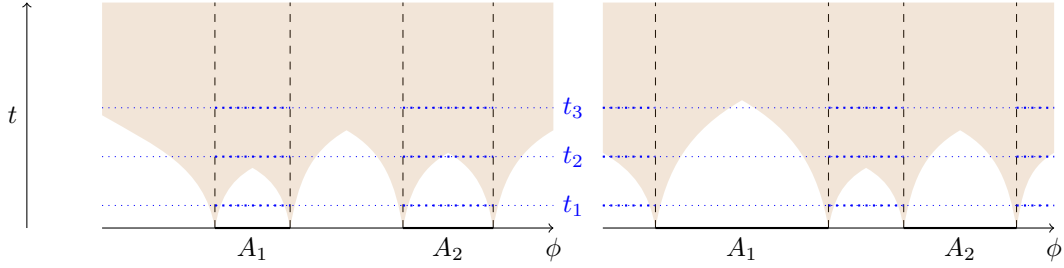
\begin{figure}[h!]
  \centering
  \begin{tikzpicture}
    \draw[->] (0,0) -- node[below,pos=1] {$\phi$} (6,0);
    \draw[thick] (1.5,0) -- node[below] {$A_{1}$} (2.5,0);
    \draw[thick] (4,0) -- node[below] {$A_{2}$} (5.2,0);
    \draw[dashed] (1.5,0) -- (1.5,3);
    \draw[dashed] (2.5,0) -- (2.5,3);
    \draw[dashed] (4,0) -- (4,3);
    \draw[dashed] (5.2,0) -- (5.2,3);
    \draw[->] (-1,0) -- node[left] {$t$} (-1,3);
    \fill[brown,opacity=.2] (1.5,0) to[out=80,in=-150] (2,.8) to[out=-30,in=100] (2.5,0) to[out=80,in=-150] (3.25,1.3) to[out=-30,in=100] (4,0) to[out=80,in=-150] (4.6,1) to[out=-30,in=100] (5.2,0) to[out=80,in=-150] (6,1.3) -- (6,3) -- (0,3) -- (0,1.5) to[out=-30,in=100] (1.5,0);
    \draw[dotted,blue,thick] (1.5,.3) -- (2.5,.3);
    \draw[dotted,blue,thick] (4,.3) -- (5.2,.3);
    \draw[dotted,blue] (0,.3) -- node[right,pos=1] {$t_{1}$} (6,.3);
    \draw[dotted,blue,thick] (1.5,.95) -- (2.5,.95);
    \draw[dotted,blue,thick] (4,.95) -- (5.2,.95);
    \draw[dotted,blue] (0,.95) -- node[right,pos=1] {$t_{2}$} (6,.95);
    \draw[dotted,blue,thick] (1.5,1.6) -- (2.5,1.6);
    \draw[dotted,blue,thick] (4,1.6) -- (5.2,1.6);
    \draw[dotted,blue] (0,1.6) -- node[right,pos=1] {$t_{3}$} (6,1.6);
  \end{tikzpicture}
  \begin{tikzpicture}
    \draw[->] (0,0) -- node[below,pos=1] {$\phi$} (6,0);
    \draw[thick] (.7,0) -- node[below] {$A_{1}$} (3,0);
    \draw[thick] (4,0) -- node[below] {$A_{2}$} (5.5,0);
    \draw[dashed] (.7,0) -- (.7,3);
    \draw[dashed] (3,0) -- (3,3);
    \draw[dashed] (4,0) -- (4,3);
    \draw[dashed] (5.5,0) -- (5.5,3);
    \fill[brown,opacity=.2] (.7,0) to[out=80,in=-150] (1.85,1.7) to[out=-30,in=100] (3,0) to[out=80,in=-150] (3.5,.8) to[out=-30,in=100] (4,0) to[out=80,in=-150] (4.75,1.3) to[out=-30,in=100] (5.5,0) to[out=80,in=-150] (6,1) -- (6,3) -- (0,3) -- (0,1) to[out=-30,in=100] (.7,0);
    \draw[dotted,blue,thick] (0,.3) -- (.7,.3);
    \draw[dotted,blue,thick] (3,.3) -- (4,.3);
    \draw[dotted,blue,thick] (5.5,.3) -- (6,.3);
    \draw[dotted,blue] (0,.3) -- (6,.3);
    \draw[dotted,blue,thick] (0,.95) -- (.7,.95);
    \draw[dotted,blue,thick] (3,.95) -- (4,.95);
    \draw[dotted,blue,thick] (5.5,.95) -- (6,.95);
    \draw[dotted,blue] (0,.95) -- (6,.95);
    \draw[dotted,blue,thick] (0,1.6) -- (.7,1.6);
    \draw[dotted,blue,thick] (3,1.6) -- (4,1.6);
    \draw[dotted,blue,thick] (5.5,1.6) -- (6,1.6);
    \draw[dotted,blue] (0,1.6) -- (6,1.6);
  \end{tikzpicture}
  \caption{Entanglement tsunami picture describing the series of HRT surfaces $X_{1} \to X_{2} \to X_{4}$ (left) and $X_{1} \to X_{3} \to X_{5}$ (right) as $t_{1} \to t_{2} \to t_{3}$.}
  \label{fig:ads-2-int-tsunami}
\end{figure}

The entanglement tsunami formula for this is
\begin{equation}
  S_{E} = \sum_{i} \frac{2c}{3} \sinh^{-1} \left\{ \frac{\beta/2\pi}{\sqrt{4\kappa \hat{\lambda}}} \sinh \frac{\frac{1}{2} \min \left[ \mathrm{vol} \left( E(t) \cap A_{i,l} \right), \mathrm{vol} \left( E(t) \cap A_{i,l}^{c} \right) \right]}{\sqrt{\left( \beta/2\pi \right)^{2} + 4\kappa\lambda}} \right\}.
  \label{eqn:ads-2-int-tsunami}
\end{equation}
We illustrate this in figure \ref{fig:ads-2-int-tsunami}.
Note the important difference with the CFT case: in the CFT case, we sum over connected components of $E(t)$ whereas here we have to sum over connected components of $A$ or $A^{c}$.

\subsection{Tsunami on the Base Space}  \label{sssec:qp-bs}
It is interesting to wonder what the entanglement velocity \eqref{eqn:v-def-qp} corresponds to on the base space, defined by the vielbein $e$ that satisfies the kernel equation of motion \eqref{eqn:kernel-eom}.
The effect of the $T \bar{T}$ deformation can be thought of as imposing state-dependent `clocks and rods' on the CFT dynamics.
We will see that the superluminality is entirely a result of the new clocks and rods.

Let the Euclidean base space be a torus of dimensions $2\pi R',\beta'$ and no rotation.
The CFT partition function on this torus can easily be calculated (assuming that the bulk dual is a Schwarzschild black hole, i.e. assuming that $\beta' < 2\pi R'$)
\begin{equation}
    \log Z [2\pi R', \beta'] = 2\pi \kappa \frac{2\pi R'}{\beta}.
    \label{eqn:BS-T2-Z}
\end{equation}
The kernel \eqref{eqn:dghc} takes the simplified form at large $c$,
\begin{align}
    Z_\lambda [2\pi R,\beta] &\approx \int dR' d\beta' e^{- \frac{\pi}{\lambda} (\beta - \beta') (R - R') + (2\pi)^2 \kappa \frac{R'}{\beta'}}.
    \label{eqn:kernel-simplified}
\end{align}
We evaluate the integral by saddle-point; the saddle-point equations obtained by varying $\beta',R'$ give
\begin{align}
  R' &= \frac{R}{2} \left\{ 1 + \frac{1}{\sqrt{1 + \frac{4 \kappa \hat{\lambda}}{(\hat{\beta}/2\pi)^2}}} \right\} \nonumber\\
    \beta' &= \frac{\beta}{2} \left\{ 1 + \sqrt{1 + \frac{4 \kappa \hat{\lambda}}{(\hat{\beta}/2\pi)^2}} \right\}.
\end{align}

The BS counterpart of the tsunami wavefront is at the same value of $t,\phi_{*}$, since this is the value of the map between the two spaces in our gauge.
However, since the metric is different, the same coordinate trajectory corresponds to a different speed,
\begin{equation}
  v_{\lambda,BS} = \frac{e_{\theta}^{2}}{e_{t}^{1}} v_{\lambda} = 1.
  \label{eqn:ads-qp-v}
\end{equation}
So, we find that the superluminal wavefront in the deformed theory is in fact just the normal luminal wavefront on the base space!
The superluminality then comes entirely from the redefinition of clocks and rods given by the topological gravity theory.

\begin{figure}[ht!]
  \centering
  \includegraphics[width=85mm]{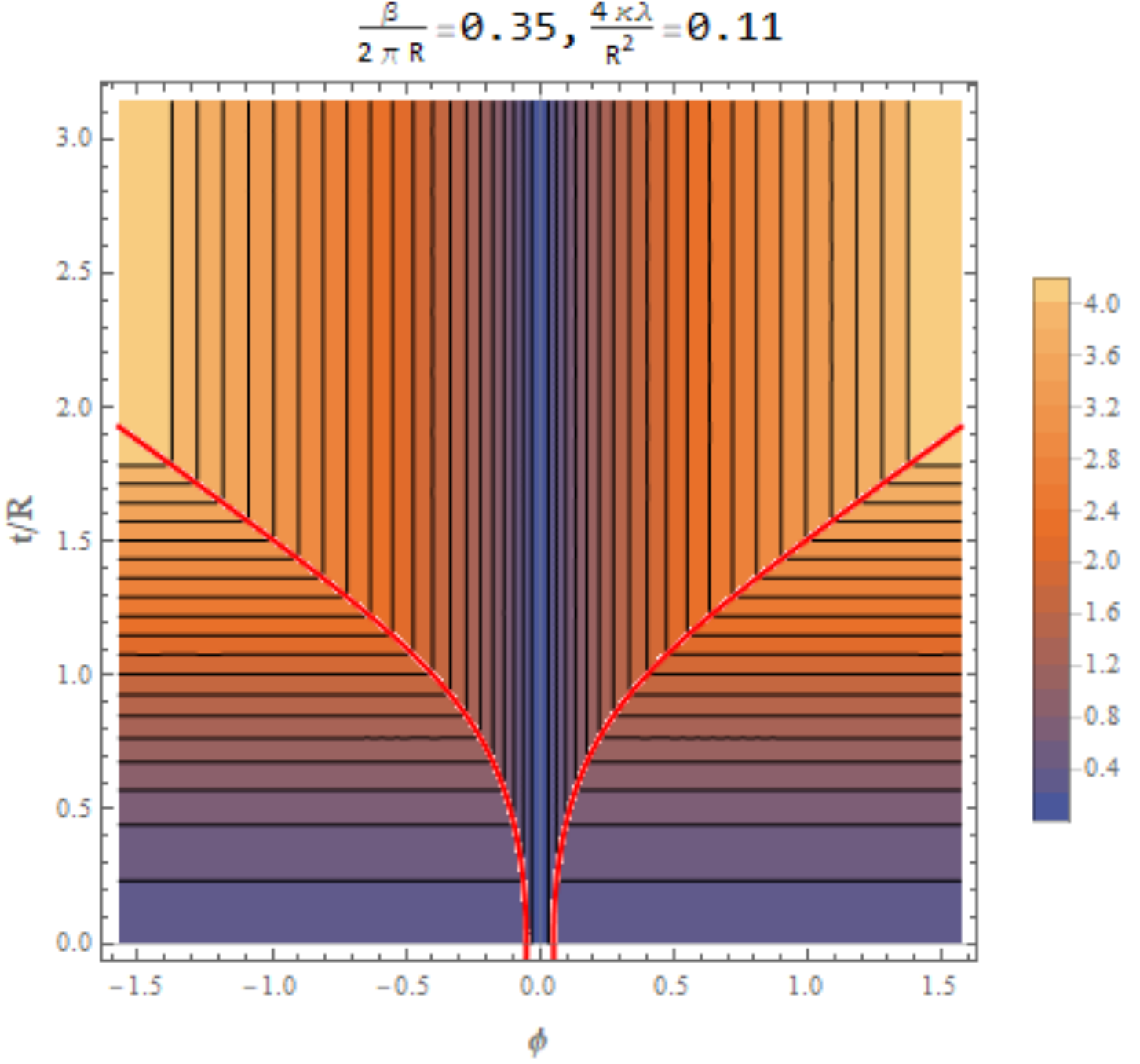}
  \caption{The entropy as a function of angular width and time. The region goes from $-\phi_{0}/2$ to $\phi_{0}/2$. At large distances, the connected HRT surface dominates and the entropy is independent of separation. At late times, the disconnected one dominates and the entropy is independent of time. The locus of transition follows \eqref{eqn:qp-traj-1}, curved at early times and superluminal after a few thermal times.}
  \label{fig:RT-bd}
\end{figure}

\section{Entanglement Tsunami in $dS_3$} \label{sec:dS_EV}
We can now repeat the calculations above for the $\Lambda_3 > 0$ setup introduced in section \ref{ssec:ds-saddle}.

\begin{figure}[ht!]
    \centering
    \includegraphics[width=.45\textwidth]{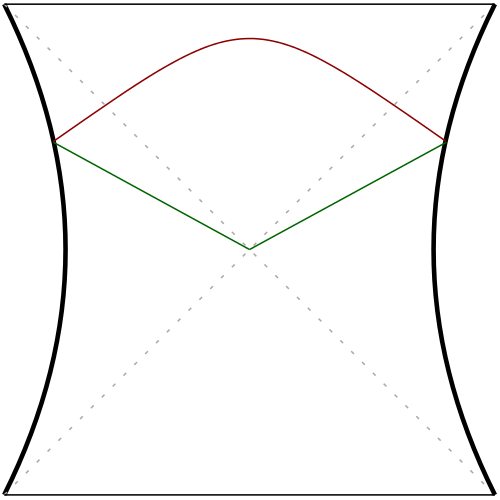}
    \hspace{.05\textwidth}
    \includegraphics[width=.45\textwidth]{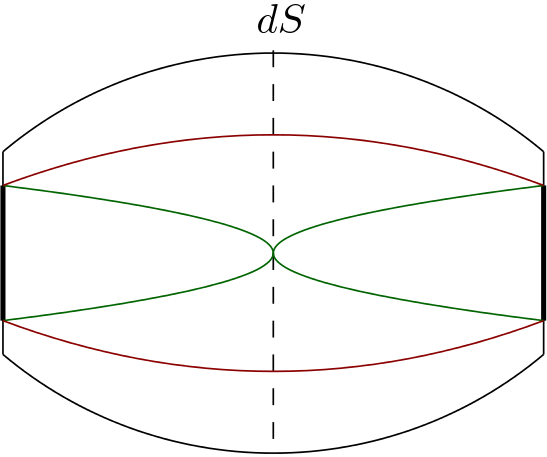}
    \caption{The two types of RT surfaces in dS. Left: Penrose diagram. Right: View from the ``top'' of the Penrose diagram.}
    \label{fig:ds-rt}
\end{figure}

In the case of one interval on each side, we again have two candidate extremal surfaces: a factorized, disconnected one and a connected, entangled one; see figure \ref{fig:ds-rt}.
Similar calculations have been done in \cite{Geng:2021wcq,Shaghoulian:2021cef,Shaghoulian:2022fop}; our prescriptions for where to put the holographic screen are different however, and that leads to some physical differences.

We begin with the factorised surface.
In static patch coordinates, the area of the disconnected surface is
\begin{align}
  A_{fact} &= 2 \ell \int_{-\phi_{0}/2}^{\phi_{0}/2} \sqrt{\frac{\rho'(\phi)^{2}}{\rho_{h}^{2} - \rho^{2}} + \frac{\rho^{2}}{\ell^{2}}} \, d\phi \equiv 2 \int_{-\phi_{0}/2}^{\phi_{0}/2} L \, d\phi.
  \label{eqn:ds-disc-1}
\end{align}
We need to extremize this area.
The translation symmetry in $\phi$ gives the `conserved quantity'
\begin{align}
  c &= \rho' \frac{\partial L}{\partial \rho'} - L = \frac{\rho^{2}/\ell^{2}}{\sqrt{\frac{\rho'^{2}}{\rho_{h}^{2} - \rho^{2}} + \frac{\rho^{2}}{\ell^{2}}}} \nonumber\\[1em]
  \Rightarrow \qquad \rho' &= \pm \frac{\rho}{\ell} \sqrt{\left( \frac{\rho^{2}}{c^{2}} - 1 \right) \left( \rho_{h}^{2} - \rho^{2} \right)}.
  \label{eqn:ds-disc-ext}
\end{align}
We see again that $c$ is the location of a turning point, and we will thus call it $\rho_{0}$ henceforth.
The sign of $\rho'$ is not fixed along the geodesic, since it has a turning point.

The solution to this equation is
\begin{align}
  \rho(\phi) &= \left[ \frac{1}{\rho_{h}^{2}} \cos^{2} \frac{\rho_{h} \phi}{\ell} + \frac{1}{\rho_{0}^{2}} \sin^{2} \frac{\rho_{h} \phi}{\ell} \right]^{- \frac{1}{2}}.
  \label{eqn:ds-disc-soln}
\end{align}
The location of the turning point is given by demanding that the surface pass through $\rho = \rho_{c}, \phi = \pm \phi_{0}/2$.
We find
\begin{align}
  \rho_{0} &= \frac{\rho_{c} \rho_{h} \sin \frac{\rho_{h} \phi_{0}}{2\ell}}{\sqrt{\rho_{h}^{2} - \rho_{c}^{2} \cos^{2} \frac{\rho_{h} \phi_{0}}{2\ell}}}, \qquad \Rightarrow \qquad \rho = \frac{\rho_{h}}{\sqrt{1 + \left( \frac{\rho_{h}^{2}}{\rho_{c}^{2}} - 1 \right) \frac{\sin^{2} \frac{\rho_{h} \phi}{\ell}}{\sin^{2} \frac{\rho_{h} \phi_{0}}{2\ell}}}}.
  \label{eqn:ds-disc-turning-pt}
\end{align}
An interesting point here is $\rho(\phi = 0) = \rho_{h}$ and so the surface always touches the horizon.
The `turning point' $\rho_{0} < \rho_{c}$ is not in the CH patch, unlike in the AdS case where $r_{h} < r_{0} < r_{c}$.

The entropy corresponding to this surface is
\begin{align}
  S_{fact} &= \frac{A_{fact}}{4 G_{N}} \nonumber\\
  &= \frac{\ell}{G_{N}} \cos^{-1} \left[ \frac{\rho_{c}}{\rho_{h}} \cos \frac{\rho_{h} \phi_{0}}{2\ell} \right] \nonumber\\
  &= \frac{2 c}{3} \cos^{-1} \left[ \sqrt{1 - \frac{\left( \hat{\beta}/2\pi \right)^{2}}{4\kappa \hat{\lambda}}} \,\cos{\left(\frac{R\phi_{0}/2}{\sqrt{4 \kappa \lambda - \left( \beta/2\pi \right)^{2}}}\right)} \right].
  \label{eqn:ds-disc}
\end{align}

\begin{figure}[ht!]
  \centering
  \includegraphics[width=100mm]{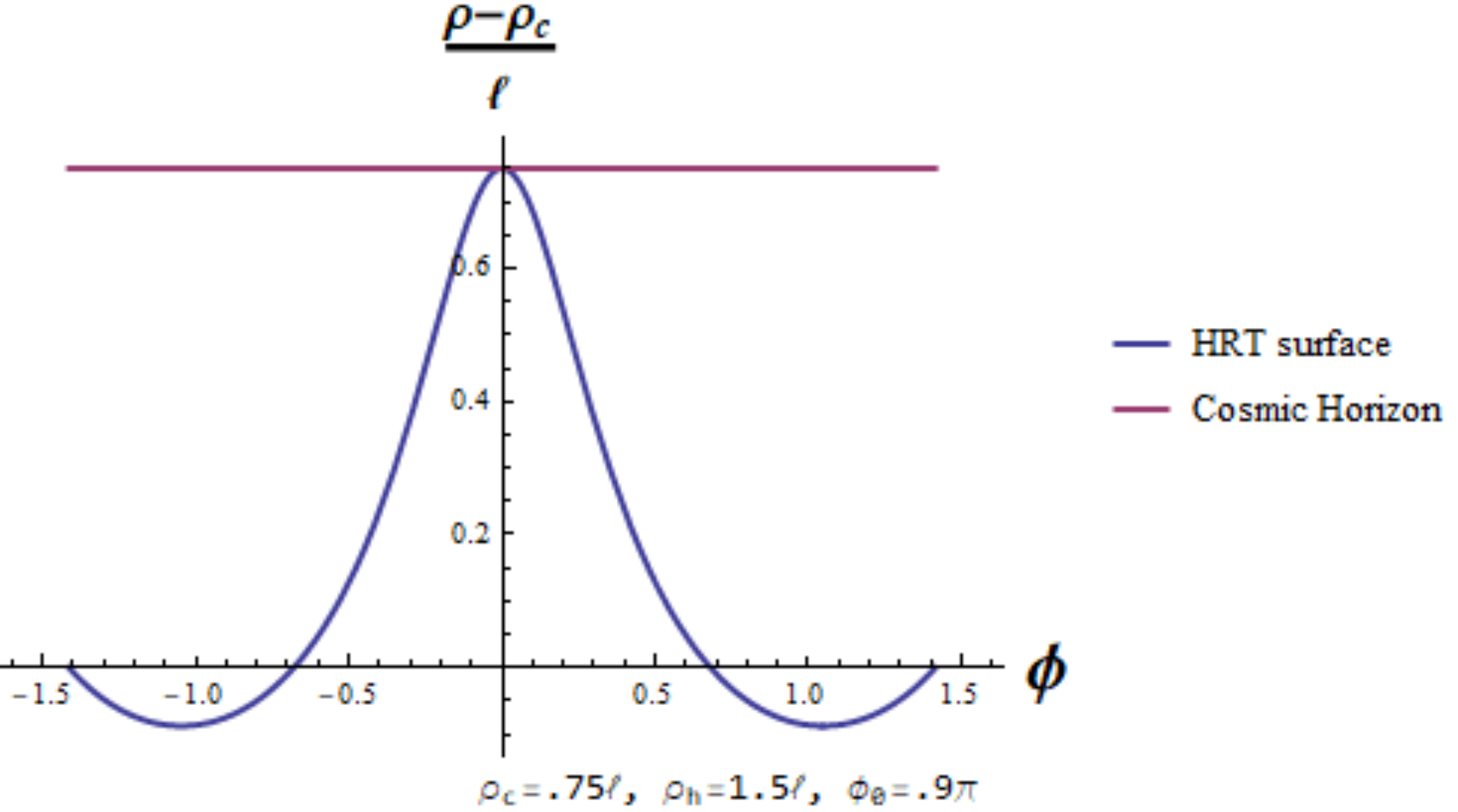}
  \caption{The disconnected HRT surface in dS when the condition \eqref{eqn:ds-disc-reality-cond} is satisfied.}
  \label{fig:ds-rt2}
\end{figure}

Since $\rho_{c} < \rho_{h}$, this entropy is necessarily real; but since $\rho_{h} \in \mathbb{R}^{+}$ and $\phi_{0} \in \left( 0,\pi \right)$, it can be negative.
More precisely, there should be an absolute value in \eqref{eqn:ds-disc}, but even then there are two cases depending on the sign of the cosine.
It is `non-positive' whenever
\begin{align}
  \frac{\rho_{h}}{\ell} \frac{\phi_{0}}{2} \ge \frac{\pi}{2},
  \label{eqn:ds-disc-reality-cond}
\end{align}
which means that the entropy becomes `negative' for \emph{some} region whenever $\rho_{h} > \ell$, i.e. when the excluded region of the 3d spacetime has a conical excess.
The source of the problem is that the RT surface passes through the excluded region $\rho < \rho_{c}$, as can be seen by the fact that in this case \eqref{eqn:ds-disc-turning-pt} is not a monotonically decreasing function in $\phi \in \left( 0, \phi_{0}/2 \right)$.
An example is illustrated in figure \ref{fig:ds-rt2}.

Interestingly, the condition \eqref{eqn:ds-disc-reality-cond} also controls the property of entanglement wedge nesting, which states that smaller regions should have smaller entanglement wedges.
For the class of regions studied here, entanglement wedge nesting is true whenever
\begin{equation}
    \left. \frac{\partial \rho(\phi)}{\partial \phi_0} \right|_\phi > 0.
    \label{eqn:ew-nesting}
\end{equation}
This simply states that the RT surface of a region of larger size is further into the bulk than that of a smaller region.
It can be easily verified from \eqref{eqn:ds-disc-turning-pt} that this is true whenever \eqref{eqn:ds-disc-reality-cond} is false.
Thus, whenever this surface makes sense it does not have the inconsistency mentioned in \cite{Shaghoulian:2021cef} in the context of a similar but different prescription to ours.

Now we turn to the connected HRT surface.
As in the AdS case, we use the Penrose diagram coordinates \eqref{eqn:ds-penrose}; the surface has two disconnected components at $\phi = \pm \phi_{0}/2$.
The area functional is
\begin{equation}
  A_{conn} = 2 \ell \int_{- y_{\partial}}^{y_{\partial}} \frac{\sqrt{1 - s'(y)^{2}}}{\cos s(y)} dy \equiv 2 \ell \int_{-y_{\partial}}^{y_{\partial}} L dy.
  \label{eqn:a-conn}
\end{equation}
Here,
\begin{align}
  \tan \frac{s_{\partial} \pm y_{\partial}}{2} &= \pm \sqrt{\frac{\rho_{h} - \rho_{c}}{\rho_{h} + \rho_{c}}} e^{\pm \frac{\rho_{h} t_{s}}{\ell^{2}}}, \nonumber\\
  \Rightarrow \qquad \tan y_{\partial} = \sqrt{\frac{\rho_{h}^{2}}{\rho_{c}^{2}} - 1} \cosh \frac{\rho_{h} t_{s}}{\ell^{2}},\ &\ \tan s_{\partial} = \sqrt{1 - \frac{\rho_{c}^{2}}{\rho_{h}^{2}}} \sinh \frac{\rho_{h} t_{s}}{\ell^{2}}.
  \label{eqn:bd-penr}
\end{align}

We again use the $y$-translation symmetry to find the HRT surface, finding the equation
\begin{equation}
  c = s' \frac{\partial L}{\partial s'} - L = - \frac{\sec s}{\sqrt{1 - s'^{2}}} \qquad s' = \pm \sqrt{1 - \frac{\sec^{2} s}{\sec^{2} s_{0}}}.
  \label{eqn:ds-conn-soln}
\end{equation}
Here, we have rewritten $c = - \sec s_{0}$, where $s' = 0$ at $s = s_{0}$.
The solution to this differential equation, demanding that $s = s_{0}$ at $y = 0$ as required by $y \to -y$ symmetry, is
\begin{equation}
  \cos y = \frac{\sin s}{\sin s_{0}}.
  \label{eqn:ds-conn-surface}
\end{equation}
Demanding that the surface passes through $\left( y_{\partial}, s_{\partial} \right)$ determines $s_{0}$ as
\begin{align}
  \sin s_{0} = \frac{\sin s_{\partial}}{\cos y_{\partial}} = \sqrt{\frac{\rho_{h}^{2}}{\rho_{c}^{2}} - 1} \,\sinh{\left(\frac{\rho_{h} t_{s}}{\ell^{2}}\right)}, \qquad \sin s = \sqrt{\frac{\rho_{h}^{2}}{\rho_{c}^{2}} - 1} \,\sinh{\left(\frac{\rho_{h} t_{s}}{\ell^{2}}\right)} \cos y.
  \label{eqn:ds-conn-final-soln}
\end{align}

It is worth noting that $s_{0}$ increases monotonically with time, till it reaches its maximum value $s_{0} = \pi/2$ when
\begin{equation}
  s_{0} = \frac{\pi}{2} \qquad \text{at} \qquad \sinh{\left(\frac{\rho_{h} t_{s}}{\ell^{2}}\right)} = \frac{1}{\sqrt{\frac{\rho_{h}^{2}}{\rho_{c}^{2}} - 1}}, \qquad \text{ i.e. } \cosh{\left(\frac{t}{\beta/2\pi}\right)} = \frac{\hat{\beta}/2\pi}{\sqrt{4\kappa \hat{\lambda}}}
  \label{eqn:ds-conn-max-time}
\end{equation}
This phenomenon was also observed in \cite{Dong:2018cuv,Narayan:2015vda}, where a specific prescription for continuing to later times was given.
We will not attempt to provide such a prescription, and will in fact find that it is a moot point.

For times before this, we can calculate the entropy of this surface (for $t_{s} > 0$) as
\begin{align}
  S_{conn} &= \frac{\ell}{G_{N}} \sin^{-1}\left[\frac{\sin y_{\partial}}{\cos s_{\partial}} \right]\nonumber\\[1em]
  &= \frac{\ell}{G_{N}} \sin^{-1} \left[ \sqrt{1 - \frac{\rho_{c}^{2}}{\rho_{h}^{2}}} \cosh{\left(\frac{\rho_{h} t}{\ell^{2}}\right)} \right] \nonumber\\[1em]
  &= \frac{2c}{3} \sin^{-1} \left[ \frac{\hat{\beta}/2\pi}{\sqrt{4\kappa \hat{\lambda}}} \cosh{\left(\frac{t}{\beta/2\pi}\right)} \right].
  \label{eqn:ds-conn-s}
\end{align}
We see that this entropy monotonically increases till the maximum time \eqref{eqn:ds-conn-max-time}, and after that it formally becomes complex.

\begin{figure}[t]
    \centering
    \includegraphics[width=0.4\linewidth]{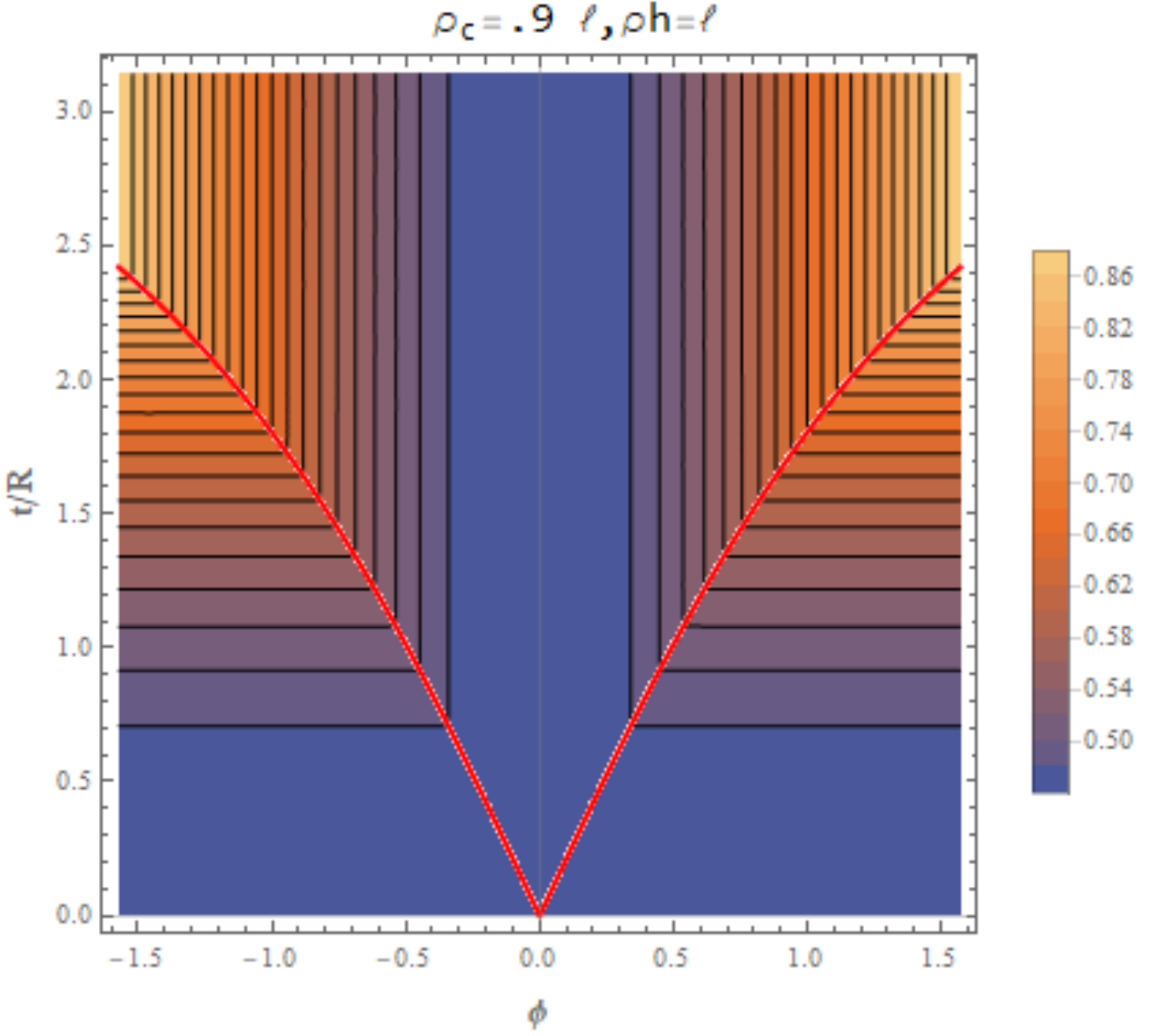} \\[0.05\linewidth]
    \includegraphics[width=0.4\linewidth]{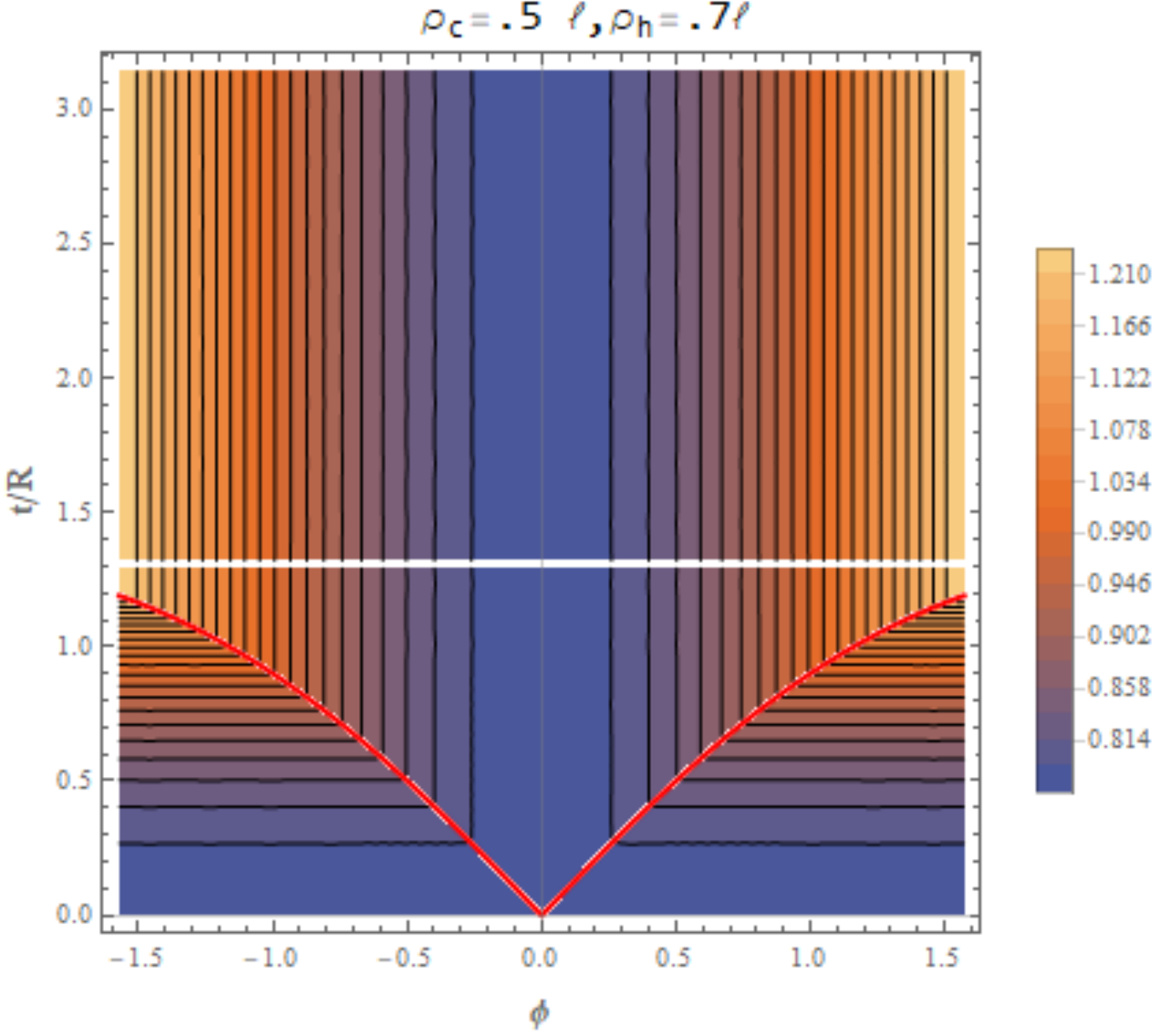} \hspace{0.05\linewidth}
    \includegraphics[width=0.4\linewidth]{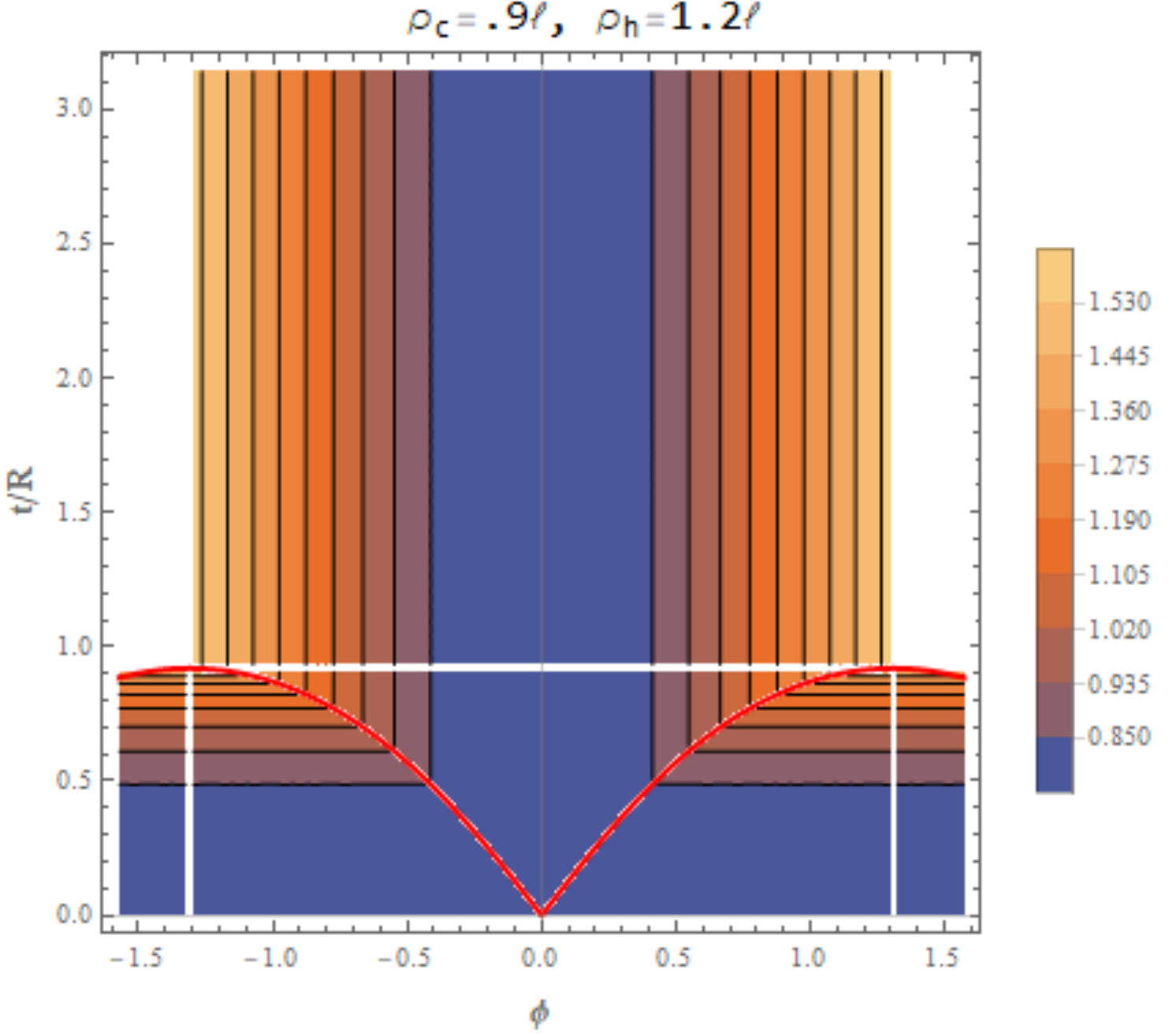}
    \caption{The entropies in the $T \bar{T} + \Lambda_{2}$ theory for the three cases of global dS${}_{3}$ ($\rho_{h} = \ell$), dS${}_{3}$ with a conical deficit ($\rho_{h} < \ell$) and dS${}_{3}$ with a conical excess ($\rho_{h} > \ell$) respectively. We have plotted the transition between the two HRT surfaces as red lines. Horizontal white lines mark the locus where the entangled surface hits future infinity, and vertical white lines mark the locus where the disconnected surface dips into the excluded region. White regions are those where neither HRT surface makes sense.}
    \label{fig:RT-dS}
\end{figure}

We now compare these two entropies and see when each one dominates.
The fact that the disconnected contribution gives a $\cos$ instead of a $\cosh$ shows that there is a fundamentally different behaviour from the case of bulk AdS.
We find it convenient to perform the comparison in terms of bulk quantities; the locus of transition is
\begin{align}
  1 - \frac{\rho_{c}^{2}}{\rho_{h}^{2}} \cos^{2}{\left(\frac{\rho_{h} \phi_{*}(t_{s})}{2\ell}\right)} &= \left[ 1 - \frac{\rho_{c}^{2}}{\rho_{h}^{2}} \right] \cosh^{2}{\left(\frac{\rho_{h} t_{s}}{\ell^{2}}\right)}.
  \label{eqn:ds-trans}
\end{align}
Some observations about this equation are
\begin{enumerate}
  \item Unlike in the AdS case, the locus of transition always intersects $\phi_{*} = t = 0$.
  \item The RHS is monotonically increasing in time, and the LHS is less than 1. This means that the late-time catastrophe \eqref{eqn:ds-conn-max-time} of the connected extremal surface is always screened off.
  \item In the limit $\rho_c \to \rho_h$, the disconnected surface dominates only for small regions $\rho_h \phi_*/\ell \sim \mathcal{O} \left(1 - \rho_c/\rho_h \right)$ or late times, $\rho_h t_s/\ell^2 \sim - \log \left(1 - \rho_c/\rho_h \right)$. The late-time effect screens off the catastrophe \eqref{eqn:ds-conn-max-time}.
  \item Similarly, for $\rho_c \to 0$, the connected surface dominates only at very early times.
  \item In the conical excess ($\rho_{h} > \ell$) case, the transition locus formally has a turning point. This is purely formal, however, since for regions larger than the turning point the factorised HRT surface does not make sense.
\end{enumerate}
We have plotted the entropies in the three cases of global dS ($\rho_{h} = \ell$), dS with a conical deficit ($\rho_{h} < \ell$) and dS with a conical excess ($\rho_{h} > \ell$) in figure \ref{fig:RT-dS}.

\subsection{Entanglement Tsunami and Multiple Intervals}
Let us now restrict to the case $\rho_{h} = \ell$, and study the tsunami picture there.
The wavefront can be found using the one-interval answer using \eqref{eqn:gen-form} and \eqref{eqn:tsunami-formula}.
We find
\begin{align}
  E_{\pm} (t) &= \left( \pm \frac{\phi_{0}}{2} - \Delta \phi(t), \pm \frac{\phi_{0}}{2} + \Delta\phi(t) \right), \nonumber\\
  &\Delta\phi (t) = \frac{1}{R} \sqrt{4\kappa\lambda - \left( \beta/2\pi \right)^{2}} \cos^{-1} \sqrt{\frac{4\kappa\lambda - \left( \beta/2\pi \right)^{2} \cosh^{2} \left( 2\pi t/\beta \right)}{4\kappa\lambda - \left( \beta/2\pi \right)^{2}}}.
  \label{eqn:ds-tsunami}
\end{align}
While the function $\Delta \phi(t)$ becomes complex at late times, the wavefronts always merge before that, as shown by the fact that the disconnected surface takes over before the connected surface hits its late-time catastrophe.\footnote{More precisely, at least one pair of wavefronts merge; but this is always the pair that is relevant for the entanglement calculation.}
The entanglement is
\begin{equation}
  S_{E} = \frac{2c}{3} \cos^{-1} \left[ \sqrt{1 - \frac{\left( \hat{\beta}/2\pi \right)^{2}}{4\kappa \hat{\lambda}}} \cos \frac{\frac{1}{4} \mathrm{vol} \left( E(t) \cap A \right)}{\sqrt{4\kappa\lambda - \left( \beta/2\pi \right)^{2}}} \right].
  \label{eqn:ds-tsunami-ee}
\end{equation}

The case of two intervals on each side also parallels the discussion in section \ref{ssec:two-int-ads} exactly.
There again five possible extremal surfaces, whose topologies agree with \eqref{eqn:ads-2-int-rts} and figure \ref{fig:ads-2-int}.
The same arguments for possible progressions among these five apply, and the tsunami formula for two intervals also has the same form as \eqref{eqn:ads-2-int-tsunami},
\begin{equation}
  S_{E} = \sum_{i} \frac{2c}{3} \cos^{-1} \left[ \sqrt{1 - \frac{\left( \hat{\beta}/2\pi \right)^{2}}{4\kappa \hat{\lambda}}} \cos \frac{\frac{1}{2} \min \left[ \mathrm{vol} \left( E(t) \cap A_{i,l} \right), \mathrm{vol} \left( E(t) \cap A_{i,l}^{c} \right) \right]}{\sqrt{4\kappa\lambda - \left( \beta/2\pi \right)^{2}}} \right].
  \label{eqn:ds-2-int-tsunami}
\end{equation}

\section{Exponential runaways in flat space} \label{sec:HPeta0}
Finally, we turn to the $\eta = 0$ case of 3d flat space.
The Rindler horizon (RH) phase still has a flat geometry, and thus all extremal surfaces are straight lines.
Thus, the calculations are very simple, and we find that the tsunami wavefront has a hyperbolic trajectory.

\begin{figure}[ht!]
    \centering
    \includegraphics[width=.20\textwidth]{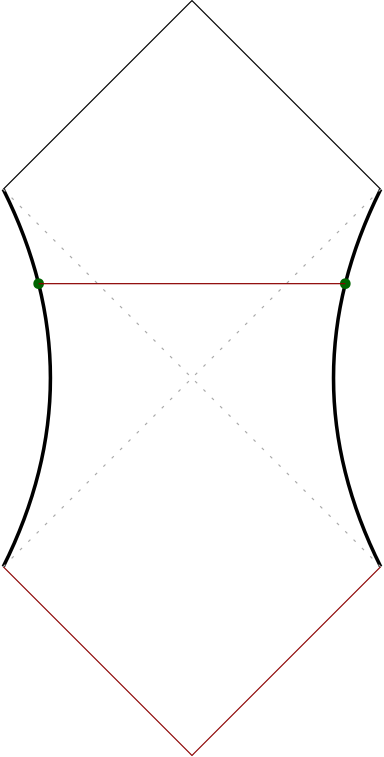}
    \hspace{.05\textwidth}
    \includegraphics[width=.45\textwidth]{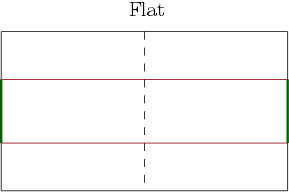}
    \caption{The two types of RT surfaces in flat space. Left: Penrose diagram. This differs from the usual Minkowski Penrose diagram in that the internal $S^1$ is the same size everywhere, including at the centre. Right: On a Cauchy slice.}
    \label{fig:flat-rt}
\end{figure}

There are two phases of the HRT surface, as before, as shown in figure \ref{fig:flat-rt}.
In the factorised phase, the HRT surface is just the sum of the lengths of the intervals,
\begin{align}
  S_{fact} &= \frac{A_{fact}}{4 G_{N}} = \frac{R \phi_{0}}{2 G_{N} \sqrt{\tilde{\lambda}}}.
  \label{eqn:flat-s-disc}
\end{align}
In the entangled phase, the surface stretches between the two boundaries.
The length of the surface can be easily calculated by transforming to Minkowski coordinates,
\begin{equation}
  T = r \sinh t_{r},\ X = r \cosh t_{r},\quad ds_{3}^{2} = - dT^{2} + dX^{2} + \frac{R^{2}}{\tilde{\lambda}} d\phi^{2}.
  \label{eqn:mink-coords}
\end{equation}
The entropy is
\begin{align}
  S_{conn} = \frac{1}{G_{N}} X\left( r = \frac{\beta/2\pi}{\sqrt{\tilde{\lambda}}}, t_{r} = \frac{t}{\beta/2\pi} \right) = \frac{\beta/2\pi}{G_{N} \sqrt{\tilde{\lambda}}} \cosh{\left(\frac{t}{\beta/2\pi}\right)}.
  \label{eqn:flat-s-conn}
\end{align}

\begin{figure}[ht!]
  \centering
  \includegraphics[width=85mm]{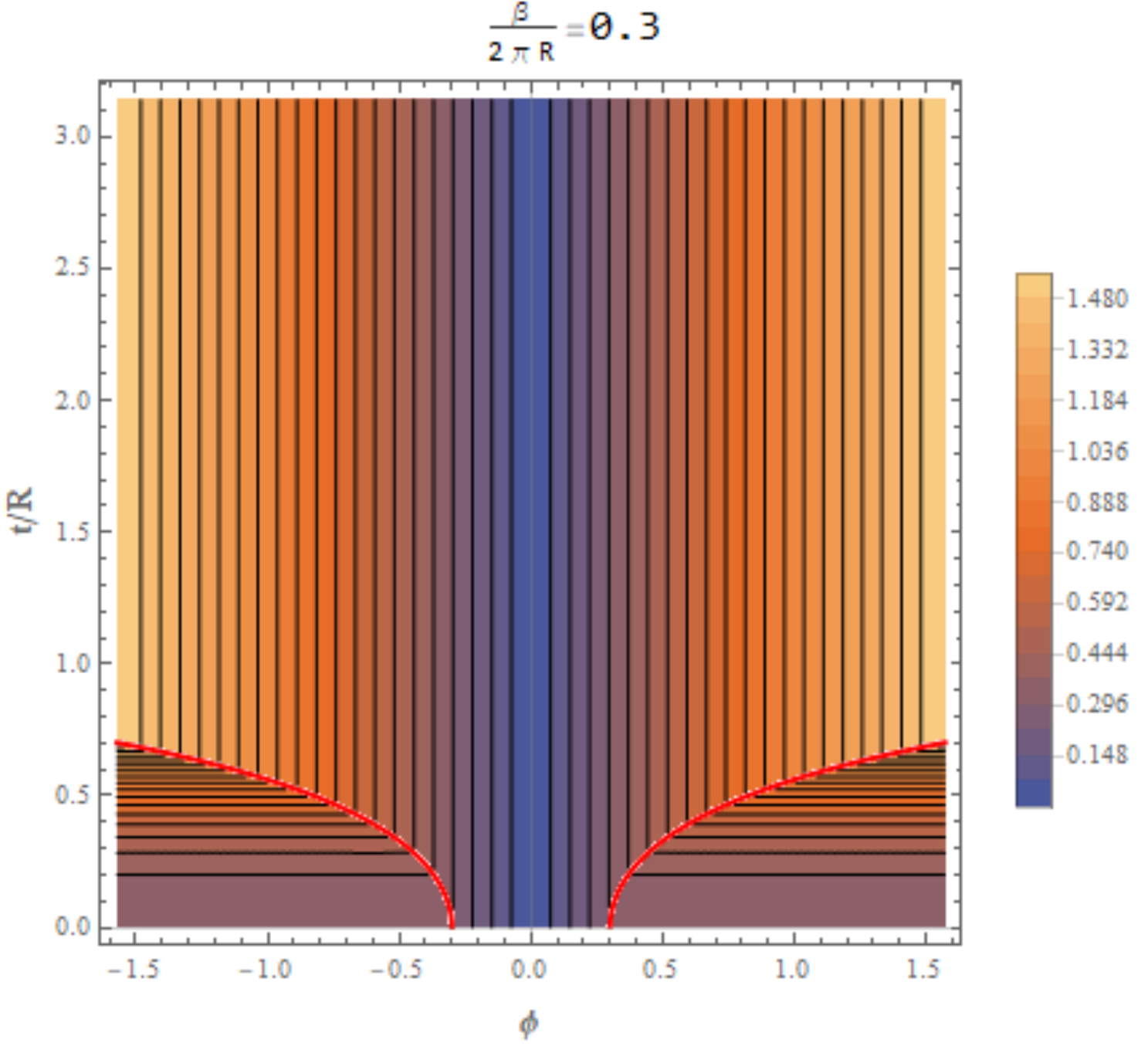}
  \caption{The entropy in units of $\frac{R}{G_{N} \sqrt{\tilde{\lambda}}}$. We see that the locus of transition grows exponentially with time.}
  \label{fig:flat-qp}
\end{figure}

The locus of transition is
\begin{align}
  \frac{R\phi_{*}}{2} = \frac{\beta}{2\pi} \cosh{\left(\frac{t}{\beta/2\pi}\right)}.
  \label{eqn:flat-transition}
\end{align}
This means that the tsunami grows exponentially with time,
\begin{align}
  E_{\pm}(t) &= \left( \pm \phi_{0} - \Delta\phi(t), \pm \phi_{0} + \Delta\phi(t) \right) \nonumber\\
  &\Delta\phi(t) = \frac{\hat{\beta}}{2\pi} \cosh \frac{t}{\beta/2\pi}.
  \label{eqn:flat-one-int-tsunami}
\end{align}
The entanglement is just
\begin{equation}
  S_{E} = \frac{1}{2} \mathrm{vol} \left( E(t) \cap A \right).
  \label{eqn:flat-one-int-ee}
\end{equation}

The case with two intervals is also straightforward.
The topologies of candidate HRT surfaces again agree with \eqref{eqn:ads-2-int-rts} and figure \ref{fig:ads-2-int}, and the same arguments about possible progressions among these with time apply.
Then, the entanglement is, similarly,
\begin{equation}
  S_{E} = 2 \sum_{i} \min \left[ \mathrm{vol} \left( E(t) \cap A_{i,l} \right), \mathrm{vol} \left( E(t) \cap A_{i,l}^{c} \right) \right].
  \label{eqn:flat-two-int-tsunami}
\end{equation}
 
\section{Discussion}
\label{sec:discussion}
In this work, we have investigated the entanglement tsunami picture of \cite{Liu:2013iza,Leichenauer:2015xra} in $T \bar{T} + \Lambda_{2}$-deformed holographic CFTs.
We have found an entanglement tsunami-type interpretation for the case of one and two symmetrically placed intervals on each side of a thermofield double.

The main open direction is to consider non-symmetric configurations and more boundary intervals and see whether our results generalise to these contexts as well.

It would be interesting to understand these results in relation to chaos and scrambling as in \cite{Mezei:2016wfz}.
It would also be intriguing to understand covariant bit threads \cite{Headrick:2022nbe} for the other two values of the cosmological constant, since that might shed some light on the local structure of the entanglement.
It is also interesting to ask if these results can be related to the failure of the split property in the deformed theory \cite{Asrat:2020uib}.

\section*{Acknowledgements}
We thank the authors of \cite{Coleman:2021nor} for discussions and collaboration on the related work \cite{Coleman:2021nor}.
This work has been partially supported by STFC consolidated grant ST/T000694/1.
RMS is supported by the Isaac Newton Trust grant ``Quantum Cosmology and Emergent Time'' and the (United States) Air Force Office of Scientific Research (AFOSR) grant ``Tensor Networks and Holographic Spacetime''.
EAC is supported by the US NSF Graduate Research Fellowship under
Grant DGE-1656518.

\bibliographystyle{JHEP}
\bibliography{refs.bib}

\providecommand{\href}[2]{#2}\begingroup\raggedright\begin{thebibliography}{10}

\bibitem{Calabrese:2009qy}
P.~Calabrese and J.~Cardy, \emph{Entanglement entropy and conformal field
  theory}, \href{https://doi.org/10.1088/1751-8113/42/50/504005}{\emph{J. Phys.
  A} {\bfseries 42} (2009) 504005}
  [\href{https://arxiv.org/abs/0905.4013}{{\ttfamily 0905.4013}}].

\bibitem{Asplund:2015eha}
C.T.~Asplund, A.~Bernamonti, F.~Galli and T.~Hartman, \emph{Entanglement
  scrambling in 2d conformal field theory},
  \href{https://doi.org/10.1007/JHEP09(2015)110}{\emph{JHEP} {\bfseries 09}
  (2015) 110} [\href{https://arxiv.org/abs/1506.03772}{{\ttfamily
  1506.03772}}].

\bibitem{Nahum:2016muy}
A.~Nahum, J.~Ruhman, S.~Vijay and J.~Haah, \emph{Quantum entanglement growth
  under random unitary dynamics},
  \href{https://doi.org/10.1103/PhysRevX.7.031016}{\emph{Phys. Rev. X}
  {\bfseries 7} (2017) 031016}
  [\href{https://arxiv.org/abs/1608.06950}{{\ttfamily 1608.06950}}].

\bibitem{Liu:2013iza}
H.~Liu and S.J.~Suh, \emph{Entanglement tsunami: Universal scaling in
  holographic thermalization},
  \href{https://doi.org/10.1103/PhysRevLett.112.011601}{\emph{Phys. Rev. Lett.}
  {\bfseries 112} (2014) 011601}
  [\href{https://arxiv.org/abs/1305.7244}{{\ttfamily 1305.7244}}].

\bibitem{Leichenauer:2015xra}
S.~Leichenauer and M.~Moosa, \emph{Entanglement tsunami in (1+1)-dimensions},
  \href{https://doi.org/10.1103/PhysRevD.92.126004}{\emph{Phys. Rev. D}
  {\bfseries 92} (2015) 126004}
  [\href{https://arxiv.org/abs/1505.04225}{{\ttfamily 1505.04225}}].

\bibitem{Coleman:2021nor}
E.~Coleman, E.A.~Mazenc, V.~Shyam, E.~Silverstein, R.M.~Soni, G.~Torroba
  et~al., \emph{{de Sitter Microstates from $T\bar T+\Lambda_2$ and the
  Hawking-Page Transition}},
  \href{https://arxiv.org/abs/2110.14670}{{\ttfamily 2110.14670}}.

\bibitem{McGough:2016lol}
L.~McGough, M.~Mezei and H.~Verlinde, \emph{{Moving the CFT into the bulk with
  $ T\overline{T} $}},
  \href{https://doi.org/10.1007/JHEP04(2018)010}{\emph{JHEP} {\bfseries 04}
  (2018) 010} [\href{https://arxiv.org/abs/1611.03470}{{\ttfamily
  1611.03470}}].

\bibitem{Zamolodchikov:2004ce}
A.B.~Zamolodchikov, \emph{{Expectation value of composite field T anti-T in
  two-dimensional quantum field theory}},
  \href{https://arxiv.org/abs/hep-th/0401146}{{\ttfamily hep-th/0401146}}.

\bibitem{Cavaglia:2016oda}
A.~Cavaglia, S.~Negro, I.M.~Szecsenyi and R.~Tateo, \emph{{$T \bar{T}$-deformed
  2D Quantum Field Theories}},
  \href{https://doi.org/10.1007/JHEP10(2016)112}{\emph{JHEP} {\bfseries 10}
  (2016) 112} [\href{https://arxiv.org/abs/1608.05534}{{\ttfamily
  1608.05534}}].

\bibitem{Cardy:2018sdv}
J.~Cardy, \emph{The $\textit{T} \overline{T}$ deformation of quantum field
  theory as random geometry},
  \href{https://doi.org/10.1007/JHEP10(2018)186}{\emph{JHEP} {\bfseries 10}
  (2018) 186} [\href{https://arxiv.org/abs/1801.06895}{{\ttfamily
  1801.06895}}].

\bibitem{GST}
V.~Gorbenko, E.~Silverstein and G.~Torroba, \emph{$\textit{dS/dS}$ and
  $\textit{T} \overline{T}$},
  \href{https://doi.org/10.1007/JHEP03(2019)085}{\emph{JHEP} {\bfseries 03}
  (2019) 085} [\href{https://arxiv.org/abs/1811.07965}{{\ttfamily
  1811.07965}}].

\bibitem{LLST}
A.~Lewkowycz, J.~Liu, E.~Silverstein and G.~Torroba, \emph{$\textit{T}
  \overline{T}$ and ee, with implications for $\textit{(A)dS}$ subregion
  encodings}, \href{https://doi.org/10.1007/JHEP04(2020)152}{\emph{JHEP}
  {\bfseries 04} (2020) 152}
  [\href{https://arxiv.org/abs/1909.13808}{{\ttfamily 1909.13808}}].

\bibitem{Tolley:2019nmm}
A.J.~Tolley, \emph{{$ T\overline{T} $ deformations, massive gravity and
  non-critical strings}},
  \href{https://doi.org/10.1007/JHEP06(2020)050}{\emph{JHEP} {\bfseries 06}
  (2020) 050} [\href{https://arxiv.org/abs/1911.06142}{{\ttfamily
  1911.06142}}].

\bibitem{Mazenc:2019cfg}
E.A.~Mazenc, V.~Shyam and R.M.~Soni, \emph{{A $\textit{T} \overline{T}$
  Deformation for Curved Spacetimes from 3d Gravity}},
  \href{https://arxiv.org/abs/1912.09179}{{\ttfamily 1912.09179}}.

\bibitem{Caputa:2020lpa}
P.~Caputa, S.~Datta, Y.~Jiang and P.~Kraus, \emph{{Geometrizing $ T\overline{T}
  $}}, \href{https://doi.org/10.1007/JHEP03(2021)140}{\emph{JHEP} {\bfseries
  03} (2021) 140} [\href{https://arxiv.org/abs/2011.04664}{{\ttfamily
  2011.04664}}].

\bibitem{Taylor:2018xcy}
M.~Taylor, \emph{{TT deformations in general dimensions}},
  \href{https://arxiv.org/abs/1805.10287}{{\ttfamily 1805.10287}}.

\bibitem{Hartman:2018tkw}
T.~Hartman, J.~Kruthoff, E.~Shaghoulian and A.~Tajdini, \emph{{Holography at
  finite cutoff with a $T^2$ deformation}},
  \href{https://arxiv.org/abs/1807.11401}{{\ttfamily 1807.11401}}.

\bibitem{Belin:2020oib}
A.~Belin, A.~Lewkowycz and G.~Sarosi, \emph{{Gravitational path integral from
  the $T^2$ deformation}},
  \href{https://doi.org/10.1007/JHEP09(2020)156}{\emph{JHEP} {\bfseries 09}
  (2020) 156} [\href{https://arxiv.org/abs/2006.01835}{{\ttfamily
  2006.01835}}].

\bibitem{Kraus:2018xrn}
P.~Kraus, J.~Liu and D.~Marolf, \emph{{Cutoff AdS$_{3}$ versus the $
  T\overline{T} $ deformation}},
  \href{https://doi.org/10.1007/JHEP07(2018)027}{\emph{JHEP} {\bfseries 07}
  (2018) 027} [\href{https://arxiv.org/abs/1801.02714}{{\ttfamily
  1801.02714}}].

\bibitem{Donnelly:2018bef}
W.~Donnelly and V.~Shyam, \emph{{Entanglement entropy and $T \overline{T}$
  deformation}},  \href{https://arxiv.org/abs/1806.07444}{{\ttfamily
  1806.07444}}.

\bibitem{Murdia:2019fax}
C.~Murdia, Y.~Nomura, P.~Rath and N.~Salzetta, \emph{{Comments on holographic
  entanglement entropy in $TT$ deformed conformal field theories}},
  \href{https://doi.org/10.1103/PhysRevD.100.026011}{\emph{Phys. Rev. D}
  {\bfseries 100} (2019) 026011}
  [\href{https://arxiv.org/abs/1904.04408}{{\ttfamily 1904.04408}}].

\bibitem{Dubovsky:2017cnj}
S.~Dubovsky, V.~Gorbenko and M.~Mirbabayi, \emph{{Asymptotic fragility, near
  AdS$_{2}$ holography and $ T\overline{T} $}},
  \href{https://doi.org/10.1007/JHEP09(2017)136}{\emph{JHEP} {\bfseries 09}
  (2017) 136} [\href{https://arxiv.org/abs/1706.06604}{{\ttfamily
  1706.06604}}].

\bibitem{Dubovsky:2018bmo}
S.~Dubovsky, V.~Gorbenko and G.~Hern\'andez-Chifflet, \emph{{$ T\overline{T} $
  partition function from topological gravity}},
  \href{https://doi.org/10.1007/JHEP09(2018)158}{\emph{JHEP} {\bfseries 09}
  (2018) 158} [\href{https://arxiv.org/abs/1805.07386}{{\ttfamily
  1805.07386}}].

\bibitem{Freidel:2008sh}
L.~Freidel, \emph{{Reconstructing AdS/CFT}},
  \href{https://arxiv.org/abs/0804.0632}{{\ttfamily 0804.0632}}.

\bibitem{Hartman:2013qma}
T.~Hartman and J.~Maldacena, \emph{Time evolution of entanglement entropy from
  black hole interiors},
  \href{https://doi.org/10.1007/JHEP05(2013)014}{\emph{JHEP} {\bfseries 05}
  (2013) 014} [\href{https://arxiv.org/abs/1303.1080}{{\ttfamily 1303.1080}}].

\bibitem{Bousso:2001mw}
R.~Bousso, A.~Maloney and A.~Strominger, \emph{{Conformal vacua and entropy in
  de Sitter space}},
  \href{https://doi.org/10.1103/PhysRevD.65.104039}{\emph{Phys. Rev. D}
  {\bfseries 65} (2002) 104039}
  [\href{https://arxiv.org/abs/hep-th/0112218}{{\ttfamily hep-th/0112218}}].

\bibitem{Goheer:2003tx}
N.~Goheer, M.~Kleban and L.~Susskind, \emph{{(1+1)-dimensional
  compactifications of string theory}},
  \href{https://doi.org/10.1103/PhysRevLett.92.191601}{\emph{Phys. Rev. Lett.}
  {\bfseries 92} (2004) 191601}
  [\href{https://arxiv.org/abs/hep-th/0310120}{{\ttfamily hep-th/0310120}}].

\bibitem{Geng:2021wcq}
H.~Geng, Y.~Nomura and H.-Y.~Sun, \emph{{Information paradox and its resolution
  in de Sitter holography}},
  \href{https://doi.org/10.1103/PhysRevD.103.126004}{\emph{Phys. Rev. D}
  {\bfseries 103} (2021) 126004}
  [\href{https://arxiv.org/abs/2103.07477}{{\ttfamily 2103.07477}}].

\bibitem{Shaghoulian:2021cef}
E.~Shaghoulian, \emph{{The central dogma and cosmological horizons}},
  \href{https://doi.org/10.1007/JHEP01(2022)132}{\emph{JHEP} {\bfseries 01}
  (2022) 132} [\href{https://arxiv.org/abs/2110.13210}{{\ttfamily
  2110.13210}}].

\bibitem{Shaghoulian:2022fop}
E.~Shaghoulian and L.~Susskind, \emph{{Entanglement in De Sitter Space}},
  \href{https://arxiv.org/abs/2201.03603}{{\ttfamily 2201.03603}}.

\bibitem{Dong:2018cuv}
X.~Dong, E.~Silverstein and G.~Torroba, \emph{De sitter holography and
  entanglement entropy},
  \href{https://doi.org/10.1007/JHEP07(2018)050}{\emph{JHEP} {\bfseries 07}
  (2018) 050} [\href{https://arxiv.org/abs/1804.08623}{{\ttfamily
  1804.08623}}].

\bibitem{Narayan:2015vda}
K.~Narayan, \emph{{Extremal surfaces in de Sitter spacetime}},
  \href{https://doi.org/10.1103/PhysRevD.91.126011}{\emph{Phys. Rev. D}
  {\bfseries 91} (2015) 126011}
  [\href{https://arxiv.org/abs/1501.03019}{{\ttfamily 1501.03019}}].

\bibitem{Mezei:2016wfz}
M.~Mezei and D.~Stanford, \emph{{On entanglement spreading in chaotic
  systems}}, \href{https://doi.org/10.1007/JHEP05(2017)065}{\emph{JHEP}
  {\bfseries 05} (2017) 065}
  [\href{https://arxiv.org/abs/1608.05101}{{\ttfamily 1608.05101}}].

\bibitem{Headrick:2022nbe}
M.~Headrick and V.E.~Hubeny, \emph{{Covariant bit threads}},
  \href{https://arxiv.org/abs/2208.10507}{{\ttfamily 2208.10507}}.

\bibitem{Asrat:2020uib}
M.~Asrat and J.~Kudler-Flam, \emph{{$T\bar{T}$, the entanglement wedge cross
  section, and the breakdown of the split property}},
  \href{https://doi.org/10.1103/PhysRevD.102.045009}{\emph{Phys. Rev. D}
  {\bfseries 102} (2020) 045009}
  [\href{https://arxiv.org/abs/2005.08972}{{\ttfamily 2005.08972}}].

\end{thebibliography}\endgroup

\end{document}